\documentclass[%
aps,
prd,
twocolumn,
nofootinbib,
amsmath,amssymb,
aps,
]{revtex4-2}
\usepackage{graphicx}
\usepackage{adjustbox}
\usepackage{dcolumn}
\usepackage{bm}
\usepackage[dvipsnames]{xcolor}
\usepackage{color, colortbl}
\usepackage[mathlines]{lineno}

\usepackage{stackengine}
\stackMath
\usepackage{xcolor}
\usepackage{dcolumn}
\usepackage{bm}
\usepackage{physics}
\usepackage{comment}
\usepackage{hyperref}

\hypersetup{
    colorlinks=true,
    menucolor = teal,
    linkcolor= olive,
    filecolor=magenta,      
    urlcolor= magenta,
    citecolor = teal, 
    pdftitle={Overleaf Example},
    pdfpagemode=FullScreen,
    }
\usepackage{float}
\usepackage{soul}

\newcommand{\rcl}{R_{\rm cl}}

\newcommand{\gpcy}{\mathrm{Gpc}^{-3} \mathrm{yr}^{-1}}

\newcommand{\osig}{\Omega_{\rm gw}^{\rm SIGW}}
\newcommand{\lnAz}{\log_{10} A_{\zeta}}

\newcommand{\lnkmn}{\log_{10} k_{\rm min} / \mathrm{Mpc}^{-1}}
\newcommand{\lnkmx}{\log_{10} k_{\rm max} / \mathrm{Mpc}^{-1}}
\newcommand{\lnrcl}{\log_{10} R_{\rm cl} / \mathrm{Gpc}^{-3} \mathrm{yr}^{-1}}

\newcommand{\mh}{M_{\rm h}}

\newcommand{\rpbh}{\rho_{\text{PBH}}}
\newcommand{\sm}{M_\odot}
\newcommand{\fpbh}{f_{\text{PBH}}}

\def\teal#1{{ \textcolor{teal}{}}}

\begin{document}

\preprint{APS/123-QED}

\title{Probing Primordial Black Hole Mergers in Clusters with Pulsar Timing Data}

\author{S\'ebastien Clesse$^{a}$}
\author{Virgile Dandoy$^{a}$}%
\author{Sonali Verma$^{a}$}%

\affiliation{${}^a$Service de Physique Th\'eorique,  Brussels Laboratory of the Universe BLU-ULB, Universit\'e Libre de Bruxelles, Boulevard du Triomphe,\\
CP225, 1050 Brussels, Belgium}

\begin{abstract}
    We consider the possibility that the stochastic gravitational wave (GW) background suggested by Pulsar Timing Array (PTA) datasets is sourced by Primordial Black Holes (PBHs). 
    Specifically, we perform a Bayesian search in the International PTA Data Release 2 (IPTA DR2) for a combined GW background arising from scalar perturbations and unresolved PBH mergers, assuming a broad PBH mass distribution. In our analysis, we incorporate constraints on the curvature power spectrum from CMB $\mu$-distortions and the overproduction of PBHs, which significantly suppress the contribution of PBH mergers to the total GW background. We find that scalar-induced GWs dominate the nHz frequency range, while PBH mergers alone cannot account for the observed signal under the standard PBH formation scenario involving Gaussian perturbations, and including only Poissonian PBH clustering. However, specific PBH models, such as those with enhanced clustering, could yield a GW background dominated by PBH mergers. Overall, we find that the IPTA DR2 strongly favors an astrophysical origin for the reported common-spectrum process over the PBH models considered in this analysis.
\end{abstract}

\maketitle

\tableofcontents

\section{Introduction}
\label{sec:level1}

Primordial Black Holes (PBHs)~\cite{1967SvA....10..602Z, 1974Natur.248...30H, Chapline:1975ojl, 1975ApJ...201....1C} might have formed in the early radiation-dominated universe from the collapse of large density perturbations upon horizon re-entry. Such large perturbations on small scales (or comoving wavenumber $k \gg 1 \ \mathrm{Mpc}^{-1}$) required for substantial PBH formation can be efficiently produced in many inflationary models (for examples see refs.~\cite{Leach:2001zf, Ballesteros:2017fsr, Mishra:2019pzq}). 
Additional PBH formation mechanisms can also be considered, for example, PBH formation can occur during a first order cosmological phase transition~\cite{Liu:2021svg, Baker:2021nyl, Kawana:2021tde, Gross:2021qgx}, or from the collapse of bubbles nucleated during inflation~\cite{Garriga:2015fdk} (see also Ref.~\cite{Carr:2020gox}). 
The existence of PBHs would thus, be connected to new physics in the early universe. 

For PBH formation from the direct collapse of primordial density perturbations, the PBH mass is roughly given by the size of the cosmological horizon at the time of the collapse. PBHs, thus, can have a wide range of masses, not limited by the Chandrasekhar limit~\cite{1931ApJ....74...81C}, unlike stellar-origin black holes.
In fact, the detection of a sub-solar mass black hole is considered to be the most robust smoking gun signature of a primordial origin.  Several candidates have recently been identified~\cite{Prunier:2023cyv,Morras:2023jvb,LIGOScientific:2022hai,Phukon:2021cus} but without enough significance to firmly claim a detection.  

PBHs may have contributed to black hole merger events detected by the LIGO-Virgo-KAGRA (LVK) Collaboration (see for e.g., Refs.~\cite{Clesse:2020ghq, DeLuca:2020sae} for PBH explanation of events GW190814~\cite{LIGOScientific:2020zkf} and GW190521~\cite{LIGOScientific:2020iuh}).
Moreover, PBHs could also explain the dark matter (DM) or a fraction of it.  These motivations have resulted in the growing interest for PBHs and ways to test these scenarios, primarily at gravitational wave experiments (see Ref.~\cite{LISACosmologyWorkingGroup:2023njw} for review).

Unresolved PBH binary mergers would also contribute to a stochastic gravitational wave background (SGWB). Furthermore, a GW background can also be sourced as a consequence of the mechanism responsible for PBH formation itself. For example, in the standard scenario for PBH formation which relies on the collapse of large primordial density perturbations, an SGWB is unavoidably induced by scalar perturbations at second order in perturbation theory~\cite{Pi:2020otn,Inomata:2019yww,Kohri:2018awv}. 

A unique probe to study the gravitational-wave background (GWB) in the nano-Hertz (nHz) frequency range is offered by Pulsar Timing Arrays (PTA). Recent results released by the NANOGrav~\cite{NANOGrav:2023hde, NANOGrav:2023gor}, EPTA (in combination with InPTA)~\cite{EPTA:2023fyk, EPTA:2023sfo}, PPTA~\cite{Reardon:2023gzh} and CPTA~\cite{Xu:2023wog} collaborations along with the joint International PTA~\cite{Antoniadis:2022pcn} provide the first convincing evidence for a stochastic GW spectrum in the nHz frequency range, with around $2-4 \ \sigma$ significance for a Hellings-Downs correlation~\cite{1983ApJ...265L..39H}. 
The origin of this GW background is still unknown, the prime explanation being a population of inspiralling supermassive black hole binaries (SMBHBs). Alternatively, 
the signal could have a cosmological origin. See for example Refs.~\cite{NANOGrav:2023hvm, Ellis:2023oxs} for a comparison between possible models.  

A PBH interpretation of the PTA signal has been widely investigated in literature~\cite{DeLuca:2020agl,  Harigaya:2023pmw, HosseiniMansoori:2023mqh, Yi:2023npi,Balaji:2023ehk, Figueroa:2023zhu, Iovino:2024tyg, Braglia:2021wwa, Ferrante:2023bgz,Madge:2023dxc,Inomata:2023zup, Dandoy:2023jot, Depta:2023qst, Gouttenoire:2023nzr}. These works have either considered only the scalar-induced GW background~\cite{DeLuca:2020agl, Harigaya:2023pmw, Ellis:2023oxs, HosseiniMansoori:2023mqh, Yi:2023npi,Balaji:2023ehk, Figueroa:2023zhu, Braglia:2021wwa, Ferrante:2023bgz,Madge:2023dxc,Inomata:2023zup, Dandoy:2023jot}, or else considered that the PTA GW background arises entirely from supermassive PBH binaries in the early Universe~\cite{Depta:2023qst, Gouttenoire:2023nzr}. 
We are not aware of any study of PTA data which considers all the relevant GW backgrounds associated to PBHs i.e. the scalar-induced GWs (SIGWs) as well as PBH mergers, for the case of broad PBH mass distributions.

In this work, we systematically address this problem and assess whether the GWB contribution from PBH binaries could be substantial enough to compete with that from SIGWs. In our analysis, instead of using a commonly assumed monochromatic or lognormal PBH mass function, we have computed the PBH merger rates for a PBH mass function starting from first principles. For a realistic phenomenological PBH model, the computation of a PBH mass function takes into account the primordial power spectrum of curvature perturbations, $\mathcal{P}_{\zeta}$, corresponding to small scales ($k \gg k_{\rm CMB} \sim 1 \ \mathrm{Mpc}^{-1}$). In addition, we have considered the effect of the lowering of the equation of state $w$, (with respect to its value in the radiation-dominated universe, $w = 1/3$)
associated to an enhancement in PBH production associated to those scales~\cite{Byrnes:2018clq} plus the effect of this variation on the collapse process itself.\\ 
For the SGWB from PBH mergers, we have considered two channels for PBH binary formation: $i)$ early PBH binaries, which form when a PBH pair decouples from the Hubble flow in the early radiation-dominated Universe, and $ii)$ late PBH binaries, which can form dynamically inside Poisson-induced PBH halos or clusters, forming right after matter-radiation equality. The merger rate from both these channels have many uncertainties as we discuss in section~\ref{Sec: GW merger}. 

A crucial aspect of our analysis is the consideration of mergers rates of dynamically formed late-time PBH binaries in halos. Poisson fluctuations~\footnote{Note that we often call this inevitable effect of Poisson fluctuations as ``clustering'', however, it is not to be confused with PBH clustering at formation which is negligible for the case in which the curvature perturbations follow Gaussian statistics~\cite{MoradinezhadDizgah:2019wjf}.} in the PBH number density will inevitably lead to the formation of small scale PBH halos or clusters with varying cluster mass. Furthermore, these PBH halos can further expand in radius with time due to gravitational heating from PBHs~\cite{Brandt:2016aco, Carr:2023tpt} with the smallest halos evaporating. In our work, we address these aspects and provide a detailed template for such merger rate computations for a broad PBH mass distribution.
While our main conclusions exclude the possibility (in line with Refs.~\cite{Gouttenoire:2023nzr, Depta:2023qst}) of explaining the PTA signal with only PBH mergers, our analysis with respect to previous works is more rigorous. Our analysis template can be useful for such searches for PBH associated GW backgrounds at other GW experiments like LVK, and future LISA.

Before our analysis, Ref.~\cite{Bagui:2021dqi} has previously considered the scenario of an SGWB in the nHz frequency range originating from PBH binaries with a broad mass distribution. As opposed to a monochromatic PBH distribution usually considered in literature, a broad PBH mass distribution features various peaks, most notably a strong peak at PBH mass $m \sim 1 \ M_{\odot}$ corresponding to the QCD epoch in the Standard Model (SM) at a temperature $T \sim 100 \ \mathrm{MeV}$~\cite{Byrnes:2018clq}. Peaks in the broad mass distribution correspond to the lowering of the equation of state $w$ of the Universe due to decoupling of SM particles. The analysis of Ref.~\cite{Bagui:2021dqi} showed that the GWB from PBH binaries forming in the late-time Universe inside PBH halos could be probed with PTA observations. This large GWB from PBH mergers was identified with a large number of merger events from PBH binaries with asymmetric masses (typically $m_1 \sim 2 \ M_{\odot}$  and  $m_2 \sim 10^2 - 10^5 \ M_{\odot}$). PBH binaries, forming in the early Universe, much before the matter-radiation equality, on the other hand,  result in a GW background that is much more peaked in the LVK frequency range corresponding to $\sim$ 10 Hz~\cite{Bagui:2021dqi}.
Contrary to the conclusions in Ref.~\cite{Bagui:2021dqi}, we find that this enhancement from asymmetric binaries is greatly reduced due to the cut-off at large scales in the primordial power spectrum from CMB $\mu$-distortions~\cite{Chluba:2012we}, which restricts the PBH mass range in merger rates to $m \lesssim 1000 \ M_{\odot}$.

In this paper, for a PBH phenomenological model with a broad mass function, we perform a Bayesian analysis of the International PTA  data release 2 (IPTA DR2)~\cite{Perera:2019sca} to investigate the possibility that the observed PTA signal can arise from a combination of SIGWs and PBH mergers. While our results show that the scalar induced GWs dominate the GWB from PBH mergers for a broad PBH mass function derived using Gaussian curvature perturbations, we also investigate the enhancement in PBH merger rates required to explain the PTA signal with only PBH mergers. For example, a much more dense PBH cluster or enhanced PBH clustering could possibly lead to much enhanced merger rates with respect to our computations. To account for this enhancement and various other uncertainties in the merger rates (discussed in Section~\ref{sec:early_binaries}), we take a phenomenological approach and treat the parameter for clustering as a free parameter and additionally perform a Bayesian search with this nuisance parameter. An enhancement of merger rates for early PBH binaries for highly clustered PBH distributions has been previously studied in Refs.~\cite{PhysRevD.99.063532, Raidal:2017mfl, Ballesteros:2018swv}. \\
As a result of these Bayesian analyses, we derive posterior probability distributions for the parameters of the primordial power spectrum, namely, the amplitude and the spectral tilt. We finally compare the predictive posterior PBH mass function derived from the Bayesian search with IPTA data in relation to other constraints on PBHs.

This paper is structured as follows: in Sec.~\ref{Sec: Mass function} we review the mechanism of PBH formation and the mass function obtained in the case of a broad primordial power spectrum of curvature perturbations. In Sec.~\ref{Sec: SIGW} and Sec.~\ref{Sec: GW merger} we review the GWB induced by scalar perturbations at second order and from PBH binaries,  respectively.  The results of the Bayesian analysis are presented in Sec.~\ref{Sec: PTAs}. In Sec.~\ref{ref:pbh_constraints}, we discuss results in relation to other PBH constraints.  We finally conclude in Sec.~\ref{Sec: Conclusion} and envisage some perspectives. We further give details of our calculations in appendices: on SIGWs (Appendix~\ref{app:sigw}), binary formation cross-section (Appendix ~\ref{app:sigma_avg}),  PBH halo formation (Appendix~\ref{app:pbh_clusters}), and related computation on the clustering factor (Appendix~\ref{App: Clustering Factor}), primordial power spectrum constraints (Appendix~\ref{App: Constraints PS}), and choice of window function for PBH abundance  (Appendix~\ref{app:window_func}). Finally, our full posterior distributions are given in Appendix~\ref{App: Posteriors}.

\section{PBH mass function from a Power Law primordial spectrum} \label{Sec: Mass function}

{In this section, we summarize the derivation of the PBH mass distribution obtained from the collapse of curvature perturbations in the Press-Schechter formalism, broadly following Ref.~\cite{Young:2019osy}}.

The standard PBH formation scenario relies on the assumption of large primordial curvature fluctuations $\zeta$ at small scales, parameterized by the curvature power spectrum $\mathcal P_{\zeta}(k)$. {At horizon crossing, a PBH is formed from the collapse of an overdensity of amplitude $\delta$ (generated by the perturbation $\zeta$) if it exceeds a critical threshold denoted by $\delta_{\rm c}$.} 

The PBH mass is then approximately close to the horizon mass at re-entry 
\begin{align}
M_{\rm H}(k) &= 4\pi M_{\rm p}^2/H \\
& \simeq 20 \ M_{\odot} \bigg(\frac{k}{10^6 \ \mathrm{Mpc}^{-1}}  \bigg)\bigg(\frac{g_{*, s}^4(T_{\rm cr}) g_{*}^{-3}(T_{\rm cr})}{17.25} \bigg)^{-1/6} 
\end{align}
with $T_{\rm cr}$ being the temperature at horizon crossing given by $k = a H(T_{\rm cr})$, and $g_*(T)$ and $g_{*,s}(T)$ are the temperature dependent effective number of relativistic degrees of freedom, contributing to the energy density and entropy density respectively.

Below we review the calculation for the PBH mass function obtained from a power law primordial spectrum of the form
\begin{equation}\label{Eq: Primoridal PS}
    \mathcal P_{\zeta}(k) = A_{\zeta} \left(\frac{k}{k_{\star}}\right)^{(n_s-1)} \Theta(k - k_{\rm min} )\Theta( k_{\rm max} -k ),
\end{equation}
where $A_{\zeta}$ is the amplitude at small scales, $k_{\star} = 10^6 \ {\rm Mpc^{-1}}$ is an arbitrary pivot scale here, fixed to correspond to $M_{H_*}  \approx 1 \, M_\odot$, $n_s$ is the spectral tilt and $k_{\rm min}$, $k_{\rm max}$ are the cut-off scales at large and small scales respectively, introduced so as to not violate the current constraints from the CMB $\mu$-distortions~\cite{Chluba:2012we}.  The higher cut-off $k_{\rm max}$ is only relevant for $n_{\rm s} \gtrsim 1$ and is introduced in order to not overproduce small-mass PBHs.   
{For broad mass distributions obtained for values of $n_{\rm s} \sim 1$, the behaviour of the equation of state of the Universe $w(T)$ with temperature, specially during the QCD epoch, can play an important role~\cite{Byrnes:2018clq,Carr:2019kxo}}.
{A non-negligible change in the effective number of degrees of freedom, $g_*(T), g_{*, s}(T)$, with temperature, leads to a transient reduction in the value of $w(T) \equiv p/\rho = 4g_{*, s}(T)/3g_*(T) - 1$, below its constant value of $1/3$ during the radiation dominated Universe. }
This translates into a simultaneous variation of the critical threshold $\delta_{\rm c}$~\cite{Byrnes:2018clq,Carr:2019kxo}. Hence, the PBH mass distribution is imprinted by the periods where $w$ and consequently the critical threshold 
decrease. This is, for instance, the case at the QCD crossover, when quarks confine into hadrons, resulting in an enhanced density of solar-mass PBHs. {For the same reason, the $W^\pm$ and $Z$ bosons decoupling would enhance the PBH production around $10^{-5} \, \rm M_{\odot}$ and the $e^+ e^-$ decoupling around $10^{6} \, \rm M_{\odot}$~\cite{Carr:2019kxo}.}\\

Under the Press-Schechter formalism of spherical collapse~\cite{Press:1973iz} and assuming Gaussian perturbations, the fraction of the radiation energy density that collapses into PBHs of mass $m$ when the scale $k^{-1}$ re-enters the horizon is given by \cite{Carr:1975qj,Gow:2020bzo}
\begin{equation}\label{eq:beta_k}
    \beta_k(m)=  \int^{\infty}_{\delta_c} {\rm d}\delta \frac{m(\delta)}{M_{\rm H}(k)} P_k(\delta) \, \delta_D\left[\ln{\frac{m}{m({\delta})}}\right].
\end{equation}
Here, $\delta$~\footnote{Note that $\delta$ here is $\delta_l$ in the notation of Ref.~\cite{Young:2019osy}.} is the smooth density contrast at linear order and $\delta_{D}$ is the Dirac delta function. The function $m(\delta)$~\cite{Choptuik:1992jv,Niemeyer:1997mt,Niemeyer:1999ak,Young:2019yug} represents the actual mass of the PBHs formed from the collapse of the overdensity $\delta$, $m(\delta) = \kappa M_{\rm H}(k)\left(\delta - \frac{1}{4\Phi} \delta^2 -\delta_c\right)^{\gamma}$ where $\gamma \simeq 0.36$ is the universal critical exponent during the radiation dominated Universe~\cite{Young:2019yug}, and $\kappa$ is a parameter depending on the particular shape of the density contrast. We set $\kappa\simeq 4$~\cite{Young:2019yug}, the typical value obtained for a nearly-scale invariant spectrum of Gaussian primordial fluctuations. The function $P_k(\delta)$ in Eq.~\eqref{eq:beta_k} is the Gaussian probability density function of the linear density contrast $\delta$, at scale $k$
\begin{equation}\label{Eq: Gaussian Distribution}
       P_k(\delta)  = \frac{1}{\sqrt{2\pi}\sigma_k} e^{-\delta^2/(2\sigma_k^2)} ~,
\end{equation}
with the variance $\sigma_k^2$ at scale $k$ given by~\cite{Gow:2020bzo, Young:2019yug}
\begin{equation}\label{eq:sigma_k}
    \sigma_{k}^2 =\frac{4}{9}\Phi^2\int_0^{\infty}\frac{\text{d}k'}{k'} \, \bigg(\frac{k'}{k} \bigg)^4 T^2(k',k) W^2(k',k) \mathcal P_{\zeta} (k')~,
\end{equation}
where $T(k',k)$ is the linear transfer function and $W(k',k)$ is the window function used to smooth the perturbations (see Appendix~\ref{app:window_func} for full equations), and $\Phi$~\cite{Franciolini:2022tfm} is a function depending on $w$ of the Universe, in the radiation dominated Universe, $\Phi = 2/3$. It is important to stress that the choice of the window function has been shown to have a strong impact on the PBH abundance~\cite{Dandoy:2023jot}~(see also the discussion in Appendix~\ref{app:window_func}). Here we choose a top-hat window function.

\begin{figure}[t!]
    \centering
    \includegraphics[width=0.5\textwidth]{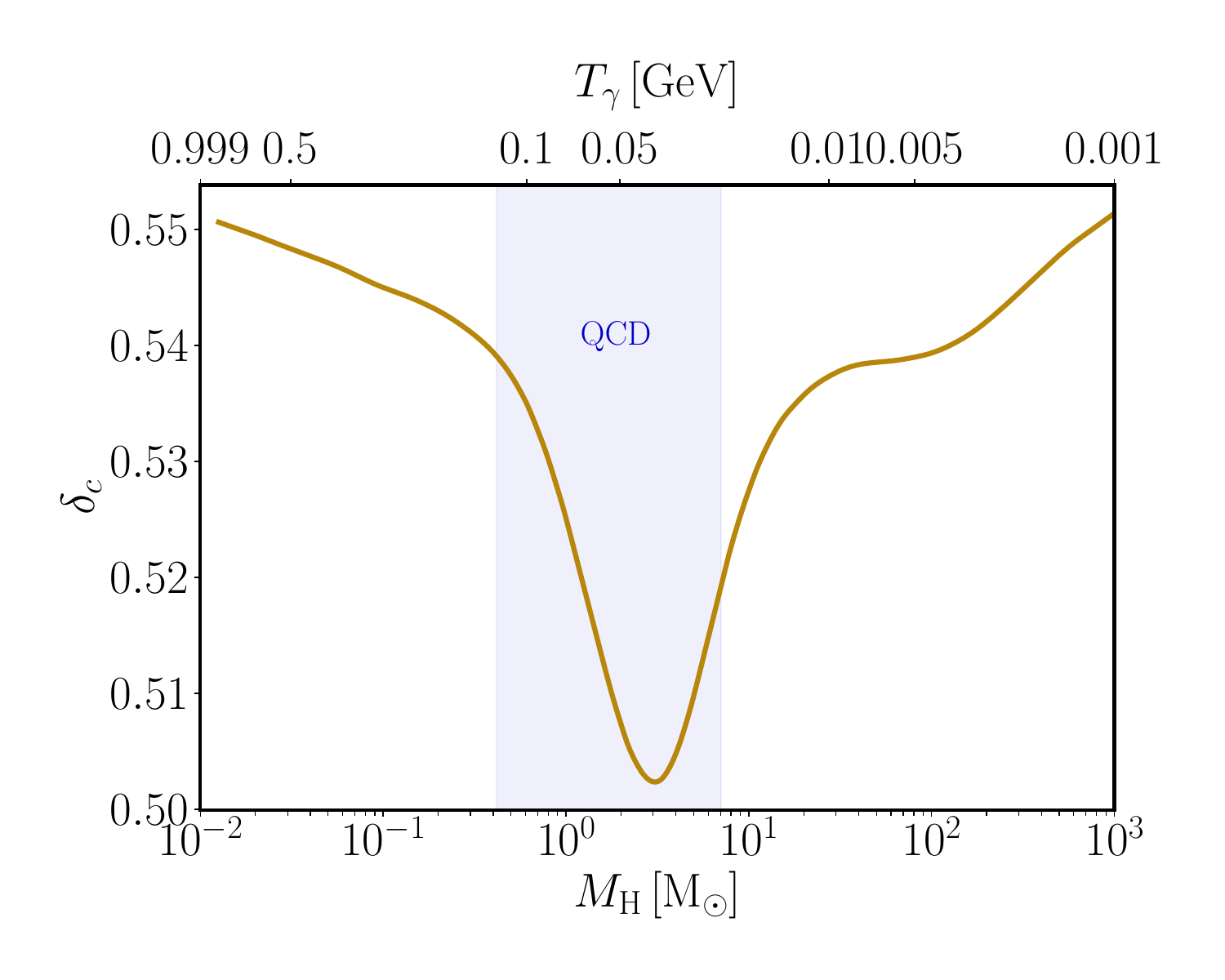}
    \includegraphics[width=0.5\textwidth]{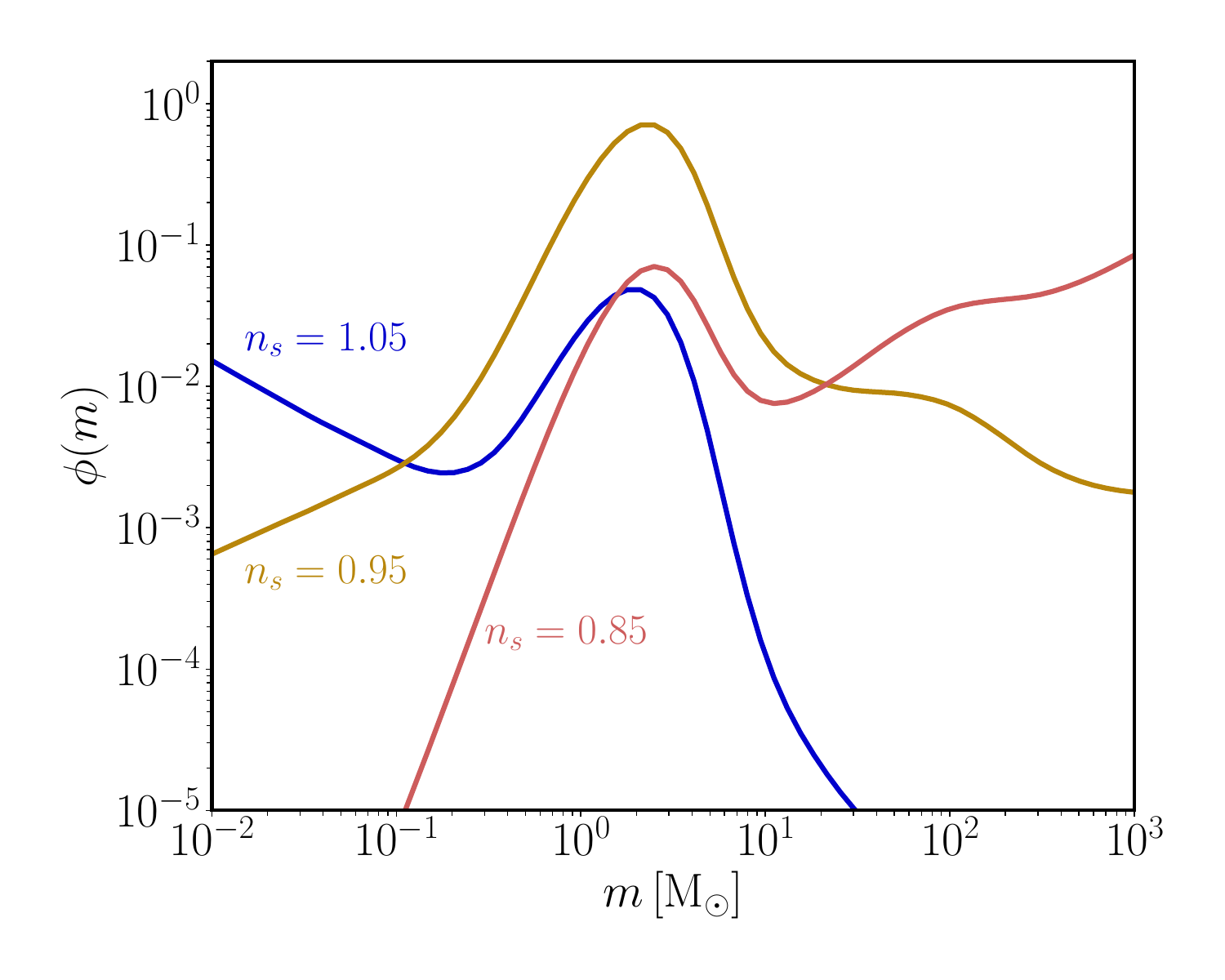}
    \caption{\textbf{Top panel:} Evolution of the critical threshold ($\delta_{\rm c}$) for PBH formation as a function of the horizon mass ($M_{\rm H})$ and the temperature of the Universe ($T_{\gamma}$). The dip around $M_{\rm H} \approx 1 M_\odot$  corresponds to the QCD phase transition. Figure adapted from Ref.~\cite{Franciolini:2022tfm}. \textbf{Bottom panel}: PBH mass function for $n_s=1.05$ (blue), $n_s=0.95$ (yellow) and $n_s=0.85$ (red). In the three cases, the amplitude of the primordial spectrum ($A_{\zeta})$ has been fixed to obtain $f_{\rm PBH} =1$ with $k_{\rm min} = 10^{4.5} \rm Mpc^{-1}$, and $k_{\rm max} = 10^{9} \rm Mpc^{-1}$.}
    \label{Fig: Critical threshold}
\end{figure}
Finally, the PBH mass function
is given by \cite{Byrnes:2018clq, Gow:2020bzo}
\begin{equation}\label{Eq: Mass function}
\begin{split}
    f_{\rm PBH}(m) &\equiv \frac{1}{\rho_{\rm DM}}\frac{{\rm d}\rho_{\rm PBH}}{{\rm d}\ln(m)}\\
    &=  \frac{2 }{ \Omega_{\rm DM}}\int \text{d}\ln k \, \left(\frac{M_{\rm eq}}{M_{\rm H}(k)}\right)^{1/2} \beta_k (m)  
    \\
    & = \frac{4 }{\gamma \,\kappa^{1/\gamma} \Omega_{\rm DM}}\int \text{d}\ln k \left(\frac{M_{\rm eq}}{M_{\rm H}(k)}\right)^{1/2}\\
    &\times \left(\frac{m}{M_{\rm H}(k)}\right)^{1+1/\gamma}\frac{ P_k(\delta(m))}{1-\frac{3}{4}\delta(m)},
\end{split}
\end{equation}
where $\Omega_{\rm DM} = 0.265$ is the dark matter relic abundance today, $M_{\rm eq} = 2.7 \times 10^{17} M_{\odot}$ {is the horizon mass at the time of matter-radiation equality} and 
\begin{equation}
\delta(m)  = 2 \Phi \ \big(1 - \sqrt{1 - \frac{1}{\Phi} ( \delta_{\rm c} + q^{1/\gamma} ) }\ \big)~,
\end{equation}
with $ q = m/\kappa M_H(k)$.   In the rest of this work, we will additionally use the normalized PBH mass function, $\phi(m) \equiv f_{\rm PBH}(m)/f_{\rm PBH} $, given by
\begin{equation}\label{Eq: Normalized distribution}
    \phi(m) = \frac{1}{{\rho}_{\rm DM} \fpbh} \frac{\text{d} \rho_{\rm PBH}}{\text{d} \ln m},
\end{equation}
with $\fpbh$ being the fraction of DM constituted by PBHs. The function $\phi(m)$ is normalized such that $\int \phi(m) \ {\rm d} \ln m = 1$.

Using this formalism, we calculate the PBH mass function for the primordial power spectrum defined in Eq.~\eqref{Eq: Primoridal PS}.
According to Refs.~\cite{Franciolini:2022tfm,Escriva:2019phb,Musco:2018rwt,Musco:2020jjb,Escriva:2022bwe}, the critical threshold $\delta_c$ must be carefully calculated as a function of the primordial power spectrum shape. In the case of the spectrum in Eq.~\eqref{Eq: Primoridal PS}, it has been shown that $\delta_c \approx 0.55$~\cite{Franciolini:2022tfm}. Moreover, we also consider the effect of the change of $w$ of the Universe on $\delta_{\rm c}$. In Fig.~\ref{Fig: Critical threshold}, we show the evolution of the critical threshold as a function of the horizon mass $M_{\rm H}$~\cite{Franciolini:2022tfm}. As already discussed, $\delta_{\rm c}$ reaches its lowest value around the QCD epoch.\\

\begin{figure}[t!]
    \centering
    \includegraphics[width=0.5\textwidth]{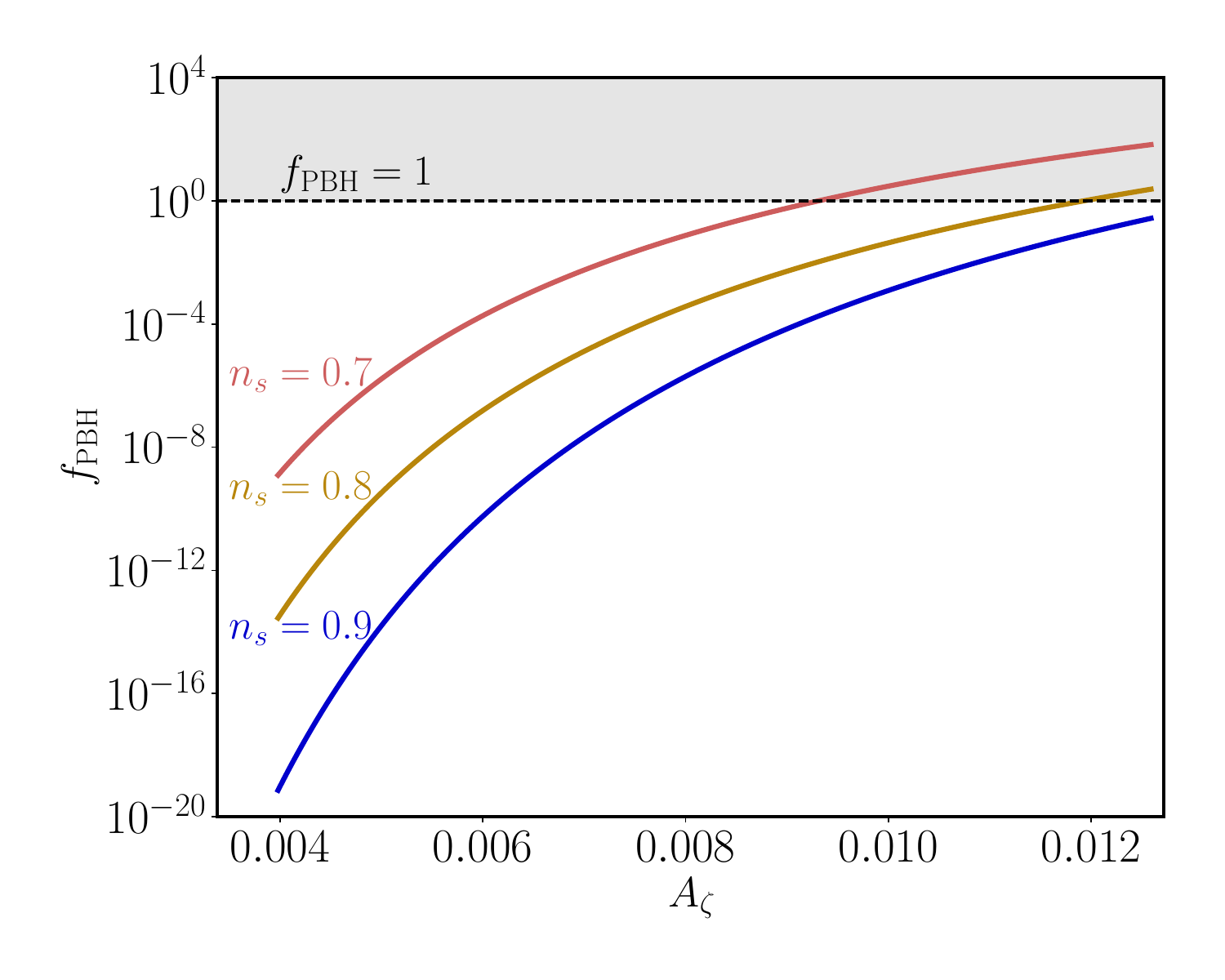}
    \caption{Total PBH abundance as a function of the amplitude $A_{\zeta}$ and the power spectral index $n_s$ of small-scale primordial fluctuations. In this example, we set $k_{\min}  = 10^{4.5} \rm Mpc^{-1}$ and $k_{\max}  = 10^{9} \rm Mpc^{-1}$. 
    }
    \label{Fig: PBH abundance }
\end{figure}

In the bottom panel of Fig.~\ref{Fig: Critical threshold} we show the corresponding mass functions for three choices of the spectral index: $n_s= 1.05$, $n_s=0.95$ and $n_s=0.85$ (for each, $k_{\min}  = 10^{4.5} \ \rm Mpc^{-1}$ and $k_{\max}  = 10^{9} \ \rm Mpc^{-1}$).
For $n_s \approx 0.95$, the mass function peaks at the solar mass level corresponding to the lowering of $\delta_{\rm c}$ during the QCD epoch. For $n_s < 0.85 $, the mass function peaks at large PBH mass, roughly corresponding to the horizon mass at $k_{\rm min}$, While for $n_s > 1.05 $, the mass function peaks at low PBH mass, roughly corresponding to the horizon mass at $k_{\rm max}$.

As already pointed out in numerous papers (see for instance Refs.~\cite{Vaskonen:2020lbd,Dandoy:2023jot}), the final PBH abundance $\fpbh$ is exponentially sensitive to the primordial power amplitude $A_{\zeta}$ (through the variance $\sigma_k^2$ defined in Eq.~\eqref{eq:sigma_k}).

This sensitivity of $\fpbh$ on the primordial spectrum parameters is illustrated in Fig.~\ref{Fig: PBH abundance }.\\


\section{Scalar Induced Gravitational Waves }\label{Sec: SIGW}

A stochastic spectrum of second-order GWs is induced by scalar perturbations (see review in Ref.~\cite{Domenech:2021ztg}).  For Gaussian perturbations, this induced GW spectrum can be sizeable for enhanced primordial curvature perturbations at small scales, typically around $\mathcal{P}_\zeta \simeq 10^{-2}$~\cite{Ananda:2006af, Bugaev_2010, Alabidi:2012ex, Baumann:2007zm}. Such large curvature perturbations are typically required for PBH formation (see Sec.~\ref{Sec: Mass function}). There are two implications for PBHs.  First, it is possible to exclude a scalar-induced origin of a GW background if the corresponding PBH abundance exceeds that of dark matter, i.e., $\fpbh > 1$~\cite{Dandoy:2023jot,Ellis:2023oxs}.  Second, it is possible to completely rule out PBH formation in certain mass ranges due to the collapse of large perturbations if no GW background is observed in the corresponding GW frequency range. For the PTA signal, it has been shown that only a small portion of the primordial power spectrum parameter space typically survives after imposing $f_{\rm PBH} <1$~\cite{Balaji:2023ehk,NANOGrav:2023hvm,Dandoy:2023jot,Ellis:2023oxs,Domenech:2024rks} unless one includes the impact of non-Gaussianity (NG) in the perturbations in computing the PBH abundance~\cite{Franciolini:2023pbf,Liu:2023ymk,Wang:2023ost,Yuan:2023ofl}. 

In the following, we summarize the calculation of SIGWs following Refs.~\cite{Pi:2020otn,Inomata:2019yww,Kohri:2018awv}. 
The observable induced GW spectrum today is defined as $\Omega_{\text{gw},0} \equiv (1/\rho_c) \  d \rho_{\text{gw}}^0/ d \, \text{ln} \, f $, where $f$ is the frequency of the GWs, $\rho_{\text{gw}}^0 $ is the energy density in GWs today, and $\rho_c$ is the critical energy density today. This can be expressed in terms of the induced GW spectrum in the radiation era through~\cite{Espinosa:2018eve, Pi:2020otn}
\begin{align}\label{eq:omegaGW_0}
    \Omega_{\text{gw}, 0}(f) \, h^2  & =  \Omega_{r}^0 h^2 \, \bigg(\frac{g_*(T_{\text{RD}})}{g_*^0} \bigg) \nonumber \\ 
    & \times \bigg( \frac{g_{*,s}^0}{g_{*,s}(T_{\text{RD}})} \bigg)^{4/3} \Omega_{\text{gw}, \, \text{RD}}(f),
\end{align}
where $\Omega_{r}^0 h^2 = 4.1 \times 10^{-5}$, is the radiation density fraction today\footnote{including photons as well as neutrinos}, while $g_*^0 = 3.36$ and $g_{*,s}^0 = 3.91$. The  last factor $\Omega_{\text{gw}, \, \text{RD}}(f)$ is the time-independent~\footnote{The time-independence of the GW spectrum results from averaging the oscillatory behaviour, deep inside the horizon~\cite{Pi:2020otn}, as detailed in Appendix~\ref{app:sigw}.} GW spectrum during the radiation dominated era (valid up to the matter-radiation equality). This can be written as a function of the curvature perturbation power spectrum~\cite{Kohri:2018awv}:
\begin{align}\label{eq:omega_ind}
    \Omega_{\text{gw}, \, \text{RD}}(k) & = \frac{1}{12} \int^{\infty}_0 {\rm d}t \int^{1}_{-1} {\rm d}s\, \Bigg[ \frac{t(2+t)(s^2-1)}{(1-s+t)(1+s+t)} \Bigg]^2 \nonumber \\ 
      &\times  \bar{I}^2(u, v) \,  \mathcal{P}_{\zeta}(u k ) \, \mathcal{P}_{\zeta}(v k),
\end{align}
with the variables $u$ and $v$ defined in terms of the integration variables $s, t$ as  $u = (s+t+1)/2$ and $v = (t-s+1)/2$. The function $\bar{I}^2(u, v)$ appearing inside the integral is defined in Appendix \ref{app:sigw} (see also Ref.~\cite{Kohri:2018awv}).   Finally, we use the following expression to relate the comoving wavenumber $k$ of the curvature mode to the frequency $f$ of the GW spectrum:
 \begin{equation}
 \begin{split}
     f &= \frac{k}{2 \pi a_0} \simeq 1.6\, \mathrm{nHz} \, \bigg( \frac{k}{10^6 \ \mathrm{Mpc^{-1}}} \bigg)\, 
 \end{split}
 \end{equation}
where $a_0=1$ is the scale factor today.

 Eq.~\eqref{eq:omega_ind} shows the dependence $\Omega_{\text{gw}} \propto \mathcal{P}_{\zeta}^2 \propto \zeta^4$. Note that Eq.~\eqref{eq:omega_ind} assumes Gaussian perturbations $\zeta$. We do not consider the impact of non-gaussianities, see Refs.~\cite{Ferrante:2023bgz,Franciolini:2023pbf,Liu:2023ymk,Wang:2023ost,Yuan:2023ofl} for the inclusion of non-Gaussianities in the SIGW calculation.
 
Considering the power spectrum in Eq.~\ref{Eq: Primoridal PS}, one can derive an approximate analytical expression for the induced GW spectrum today~\cite{Kohri:2018awv}, given by

\begin{align}\label{Eq: Analytical Omega_gw}
    \Omega_{\rm gw, 0}^{\rm SIGW} (f) \, h^2 & = 7.5 \times 10^{-5} g_*(T) \, (g_{*,s}(T))^{-4/3} \ Q(n_s) \\ \nonumber
    & \times A_{\zeta}^2\left(\frac{f}{f_\star}\right)^{2\left(n_s-1\right)} \Theta(f - f_{\rm min} )\Theta( f_{\rm max} - f ),
 \end{align}
where $Q(n_s)$ can be computed by integrating Eq.~\eqref{eq:omega_ind} and is an $\mathcal{O}(1)$ factor~\cite{Kohri:2018awv}. For example, for $n_s = 0.9655$, one gets $Q(n_s) = 0.814$. Here, $f_{\rm min} \ (f_{\rm max})$ correspond to the $k_{\rm min} \ (k_{\rm max})$ cut-offs in the frequency space.\\
In the case of the spectrum defined in Eq.~\eqref{Eq: Primoridal PS} because of the large (small)-scale cutoff $k_{\rm min}$($k_{\rm max}$), one has to compute the SIGW spectrum numerically directly using Eq.~\eqref{eq:omega_ind}. For our analysis, we used the approximate solution above for reducing the computational time.


\section{GW background from PBH binaries}\label{Sec: GW merger}

 In this section, we determine the GW spectrum from PBH binary mergers. Two main channels are commonly considered. The first one comes from early Universe PBH binaries that decoupled from the Hubble flow before matter-radiation equality ~\cite{Nakamura:1997sm,Ioka:1998nz} (see also Refs.~\cite{Raidal:2018bbj,Vaskonen:2019jpv} for detailed computations). The second channel instead relies on PBH binaries that can dynamically form in clusters via capture.  The formation of such small-scale PBH clusters is inevitable due to the Poisson fluctuations in the spatial distribution of PBHs at formation~\cite{Clesse:2020ghq,Hutsi:2019hlw,Carr:2023tpt,Meszaros:1975ef} (details on PBH clustering can be found in Appendix~\ref{app:pbh_clusters}). The formation of these small-scale PBH clusters can be described using the Press-Schechter formalism~\cite{Press:1973iz} (see Appendix~\ref{app:pbh_clusters}). N-body simulations of Ref.~\cite{Inman:2019wvr} (and more recently done in Ref.~\cite{Delos:2024poq}) further confirm this~\footnote{These PBH clusters could even correspond to ultra-faint dwarf galaxies~\cite{Clesse:2017bsw}.}. 

In Ref.~\cite{Clesse:2020ghq}, it was shown that for broad mass spectra (which can be obtained from the PBH mass function defined in Eq.~\ref{Eq: Mass function} of Sec.~\ref{Sec: Mass function} for values of $n_{\rm s} \simeq 1$), the GW background from late-time PBH binaries in dense clusters would be enhanced compared to the case of a monochromatic or lognormal mass function. This enhanced GW amplitude was identified in Ref.~\cite{Bagui:2021dqi} as originating from the numerous binaries with asymmetric masses. Furthermore, it has been shown in Ref.~\cite{Bagui:2021dqi} that for $n_{\rm s}$ close to $0.95$, the GW background from clustered PBHs could reach the NANOGrav sensitivity and largely dominate over the early universe binaries which instead, dominate the GW background in the LVK frequency range corresponding to $10 \ \mathrm{Hz} - 10 \ \mathrm{kHz} $. We note here that the conclusions of Ref.~\cite{Bagui:2021dqi} were derived for a small-scale primordial spectrum without any cut-off at small wavenumber $k_{\rm min}$. Constraints from $\mu-$distortions (see discussion in  appendix~\ref{app:firas}) from COBE/FIRAS~\cite{Mather:1993ij, Fixsen:1996nj} which require $k_{\rm min} \gtrsim 10^5 \ \mathrm{Mpc}^{-1}$, will limit the formation of any PBHs above $m \gtrsim 1000 \ M_{\odot}$. In this scenario, we find that the GWB expected from late binaries will be much smaller than the one calculated in Ref.~\cite{Bagui:2021dqi}.

In the following subsections, we review the calculation of the GW background from PBH binaries and then discuss the merger rates for early PBH binaries and for late PBH binaries formed in small-scale clusters. \\

\subsection{PBH merger rate from early binaries}
\label{sec:early_binaries}
As we indicated earlier, an important binary formation channel comes from PBH pairs that decoupled from the Hubble flow and formed binaries before the matter-radiation equality. The merger rate of such PBH binaries formed in the early Universe is~\cite{Raidal:2018bbj}
\begin{align}\label{eq:early_merger}
    \frac{{\rm d} \mathcal{V}(z)_{\rm early}}{{\rm d} \ln m_1 {\rm d} \ln m_2} &= \frac{1.6 \times 10^6}{\rm Gpc^3 \ yr} \ \fpbh^{53/37} \bigg(\frac{t}{t_0} \bigg)^{-\frac{34}{37}} \bigg(\frac{M}{M_{\odot}} \bigg)^{-\frac{32}{37}}  \\[0.1cm] \nonumber
    & \times \eta^{-\frac{34}{37}} \ S(\phi, \fpbh, M) \ \phi(m_1) \ \phi(m_2)
\end{align}
where $M = m_1 + m_2$, $\eta = m_1 m_2/M^2$, and $t_0$ is the age of the universe. The suppression factor $S(\phi, \fpbh, M) < 1$ accounts for effects in the early and late-time Universe, $S \equiv S_{\rm early} \times S_{\rm late}$. The suppression factor from the early Universe originates due to the interaction of PBH binaries with a third PBH as well as surrounding matter perturbations, that can disrupt the binary. This is given by (see for e.g. Ref.~\cite{Raidal:2018bbj, Hutsi:2020sol})
\begin{equation}
    S_{\rm early} \approx 1.42 \bigg( \frac{ \langle m^2 \rangle / \langle m \rangle^2}{\bar{N} + C} + \frac{\sigma_{\rm M}^2}{\fpbh^2} \bigg)^{-21/74} \exp(-\bar{N})
\end{equation}
where $\sigma_{\rm M} \simeq 0.004$, C is a fitting function that we take from Appendix A of Ref.~\cite{Hutsi:2020sol} while $\bar{N}$ is the expected number of PBHs that fall into the PBH binary~\footnote{Note that the averaged mass-dependent quantities above are defined in our notation as:
\begin{equation}\label{eq:m_avg_vask}
    \langle X \rangle = \frac{\int X m^{-1} \phi(m) {\rm d} \ln m}{\int m^{-1} \phi(m) {\rm d} \ln m}
\end{equation}
}
\begin{equation}
    \bar{N} \approx \frac{M}{\langle m \rangle} \frac{ \fpbh}{\fpbh + \sigma_{\rm M}}.
\end{equation}
PBH mass distributions with a small $\langle m \rangle$ will result in $\bar{N} \gg 1$, giving an exponential suppression in the merger rate from early binaries. 

The suppression in the early merger rate in the late Universe denoted by $S_{\rm late}$ occurs due to the absorption of early PBH binaries by small-scale PBH clusters (see for example Ref.~\cite{Vaskonen:2019jpv}). The suppression factor at any redshift $z$ is accounted for by the probability of finding unperturbed PBHs outside of unstable PBH clusters. We use the following numerical fit for this suppression factor given in Ref.~\cite{Hutsi:2020sol}
\begin{equation}\label{eq:S_late}
    S_{\rm late}(z = 0) \approx \mathrm{min} \big[ 1, 9.6 \cdot 10^{-3} x^{-0.65} \exp(0.03 \, \mathrm{ln}^2 x) \big]
\end{equation}
where $x =  (t(z)/t_0)^{0.44} \fpbh$. It is important to note that the late-time suppression factor above has been computed for a monochromatic mass distribution~\cite{Vaskonen:2019jpv}. The extension for wider mass PBH functions is non-trivial and still an open issue. We will extrapolate the result for the narrow mass distribution of Eq.~\eqref{eq:S_late} also for broad mass spectra. Note that for small $\fpbh \simeq 0.004$~\cite{Vaskonen:2019jpv}, $S_{\rm late}(z = 0) \simeq 1$ for the monochromatic PBH mass distribution case, implying no suppression in the early merger rate from PBH cluster effect.  \\[0.1cm]

\subsection{PBH merger rate in late-Universe halos}
\label{sec:late_binaries}
After the epoch of recombination in the late Universe, PBH binary formation can also take place dynamically via two-body capture. A crucial ingredient for this channel is the formation of PBH clusters due to the enhanced matter power spectrum at small scales, inevitably induced by the Poisson fluctuations in the PBH number density (see Refs.~\cite{Inman:2019wvr, Meszaros:1975ef, Kadota:2020ahr}). This small-scale structure formation can be described using the Press-Schechter formalism (more details on PBH cluster formation can be found in Appendix~\ref{app:pbh_clusters}). After the matter-radiation equality, these PBH clusters can form and virialize.  In this section, we give details of the merger rates from PBH binaries that can form dynamically in such dense PBH clusters.
 
 We first proceed with the calculation of the merger rate in a single PBH halo, denoted by $\mathcal{R}_{\rm h}$. Let us consider two PBHs with masses $m_1$ and $m_2$ moving with a relative velocity $v$ in a PBH halo of mass $M_{\rm h}$. If the energy loss due to gravitational waves exceeds the orbital kinetic energy, the objects form a binary system.
The cross-section for this binary formation process is given by~\cite{Mouri:2002mc, 1989ApJ...343..725Q}
\begin{equation}\label{Eq: cross section}
    \sigma_{\rm bin} = 2 \pi \bigg(\frac{85 \pi}{6 \sqrt{2}} \bigg)^{2/7} \frac{G^2 (m_1 + m_2)^{10/7} (m_1 m_2)^{2/7}}{c^{10/7} \ v^{18/7}} .
\end{equation}\\
 The differential merger rate per unit logarithmic mass is then given by~\cite{Bird:2016dcv}:
\begin{equation}\label{eq:R_h}
\begin{split}
    \frac{{\rm d} \mathcal{R}_{\rm h}}{{\rm d} \ln m_1 {\rm d} \ln m_2 } & = \frac{{4} \, \pi}{m_1 m_2} \int^{r_{\rm h}}_0  \text{d} r \ r^2 {\frac{1}{2}} \bigg(\frac{{\rm d} \rpbh(r)}{{\rm d} \ln  m_1} \bigg)\\
    & \times \bigg( \frac{{\rm d} \rpbh(r)}{ {\rm d} \ln  m_2} \bigg) \, \langle \sigma_{\rm bin} \ v \rangle \, .
\end{split}
    \end{equation}
Here the quantity $\langle \sigma_{\rm bin} \,  v \rangle$ denotes the thermal average of the binary formation cross-section defined in Eq.~\eqref{Eq: cross section}, whose detailed calculation can be found in Appendix~\ref{app:sigma_avg}. The local PBH density distribution in the halo is captured by ${\rm d} \rpbh(r)/ {\rm d} \log m = \rho_{\rm NFW}(r) f_{\rm PBH}\phi(m)$, where $\phi(m)$ is the PBH mass distribution defined in Eq.~\eqref{Eq: Normalized distribution} and $\rho_{\rm NFW}(r)$ is the density profile of the cluster. {We have assumed a Navarro-Frenk-White (NFW) profile, $\rho_{\rm NFW}(r) = \rho_{\rm s} [ (r/r_{\rm s})(1 + \, r/r_{\rm s})^2) ]^{-1
}$~\cite{Navarro:1995iw}, with a characteristic radius $r_{\rm s}$ and characteristic density $\rho_{\rm s}$.  {$r_{\rm h}$ is the Virial radius of the cluster, defined as the radius where the halo density is equal to 200 times the critical cosmological density. They are related by a concentration parameter $C$, defined as $C \equiv r_{\rm h} / r_{\rm s}$. The density profile of halos is fully determined once $C$ has been expressed as a function of the halo mass $M_{\rm h}$. Following Ref.~\cite{Bird:2016dcv}, we use the concentration-mass relations from Refs.~\cite{Ludlow:2016ifl,Prada:2011jf}}. The total mass of the halo within $r_{\rm h}$ is given by $M_{\rm h} = 4\pi \, \rho_{\rm s} \, r_{\rm s}^3 \, g(C) $ where $g(C) = \ln(1+C) -C/(1+C)$.  

Finally, the total cosmological merger rate of PBHs can be computed by convoluting the merger rate in a single halo $\mathcal{R}_{\rm h}$ with the halo mass function ${\rm d} n/ d M_{\rm h},$~\cite{Bird:2016dcv}
\begin{equation}\label{eq:V_z}
    \frac{{\rm d} \mathcal{V}(z)}{{\rm d} \ln m_1 {\rm d} \ln m_2} = \int_{M_{\rm min}(z)}^{\infty} \text{d} M_c \frac{{\rm d} \mathcal{R}}{{\rm d} \ln m_1 {\rm d} \ln m_2 } \frac{{\rm d} n(z)}{{\rm d} M_{\rm h}} ,
\end{equation}
where the halo mass function above includes haloes from Poisson fluctuations arising from the discrete nature of PBHs (see similar calculations in Refs.~\cite{Ali-Haimoud:2017rtz, DeLuca:2020jug}). We use the PBH halo mass function derived using the Press-Schechter formalism in Ref.~\cite{Inman:2019wvr}}

\begin{align}\label{eq:PS_haloMF}
    \frac{{\rm d} n(z)}{{\rm d} \mh} =  \frac{\bar{\rho}_{\text{PBH}}} {\sqrt{\pi}} \bigg(\frac{M_{\rm h}}{M_{*}(z)} \bigg)^{1/2} \frac{e^{-M_{\rm h}/ M_{*}}}{M_{\rm h}^2}
\end{align}
with $\bar{\rho}_{\text{PBH}}$ being the background PBH density {today} given by $\bar{\rho}_{\text{PBH}} = \fpbh \rho_{\rm DM}$ and $M_{*}$ being the characteristic halo mass formed at redshift $z$~\cite{Inman:2019wvr} (see also Eq.~\eqref{eq:Mstar_full} in appendix~\ref{app:pbh_clusters})

\begin{equation}\label{eq:Mstar_main}
    M_*(z) \simeq \bigg(\frac{3656}{1 + z} \bigg)^2  \langle \fpbh^2  m \rangle.
\end{equation}
Compared to~\cite{Hutsi:2019hlw} that applied this to monochromatic or peaked mass distributions, we have averaged $\fpbh^2 m$ for the considered broad mass functions, replacing $\fpbh^2 m$ with $\langle \fpbh^2  m \rangle$ (see Eq.~\eqref{eq:fsq_mavg} for the definition of the averaged quantity in this case). 
Due to the exponential cutoff in ${\rm d}n/{\rm d}M_{\rm h}$ at large halo masses, the upper limit of the integral in Eq.~\eqref{eq:V_z} does not impact the final merger rate. Indeed, the biggest contribution to the merger rate will come from the lightest PBH halos.

The lower limit of the integral in Eq.~\eqref{eq:V_z}, thus, must be carefully considered: once formed, PBH halos can undergo dynamical heating and therefore, expand. Eventually, the small halos are completely diluted in larger halos~\cite{Carr:2023tpt} and do not contribute to the merger rate in Eq.~\eqref{eq:V_z}.  We parameterize this effect with a lower halo mass limit $M_{\rm min}(z)$ that encodes the lowest cluster mass that has not been diluted.   The calculation of $M_{\rm min}(z)$ is detailed in Appendix~\ref{App: Clustering Factor}.

\begin{figure}[t!]
    \centering
    \includegraphics[width=0.5\textwidth]{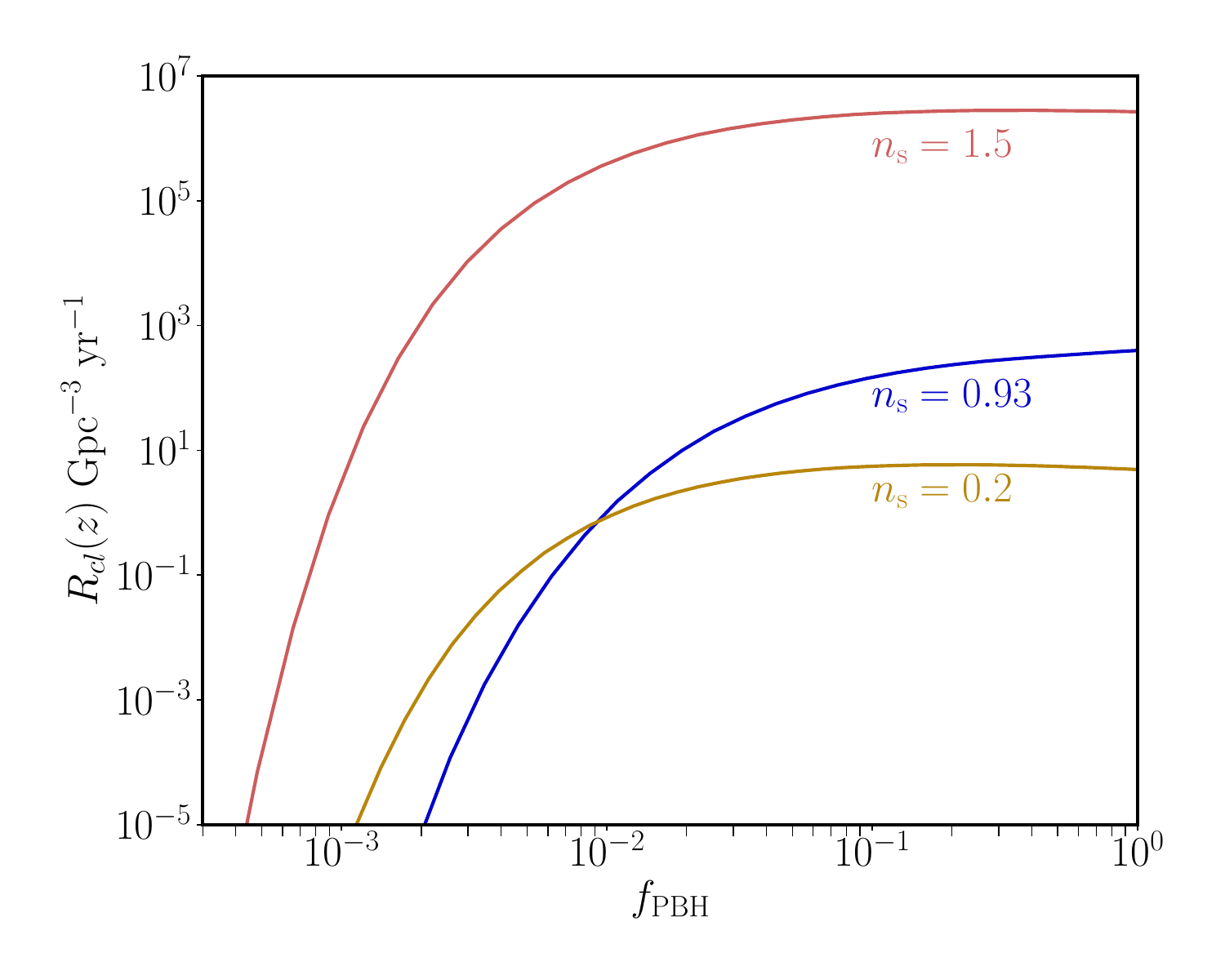}
    \caption{The clustering parameter or $R_{\rm cl}$ appearing in the merger rate for late binaries in Eq.~\eqref{Eq:Merger Rate} (see Eq.~\eqref{eq:rcl_defn} in appendix for exact expression) as a function of $\fpbh$, evaluated at redshift $z = 0$ (today), for three different values of spectral index: $n_{\rm s} = 1.5$ (red), $n_{\rm s} = 0.93$ (blue), $n_{\rm s} = 0.2$ (yellow).
    }
    \label{Fig:rcl_fpbh}
\end{figure}

Using Eqs.~\eqref{Eq: cross section},\eqref{eq:R_h}, \eqref{Eq: Normalized distribution}, we can now write the differential total merger rate per unit logarithmic mass, as a function of redshift
\begin{align}\label{Eq:Merger Rate}
    \frac{{\rm d} \mathcal{V}(z)_{\rm late}}{{\rm d} \ln m_1 {\rm d} \ln m_2} &= \fpbh^2  R_{\text{cl}}(z) \;  \phi(m_1) \; \phi(m_2) \nonumber \\
    &\times \frac{(m_1 + m_2)^{10/7}}{(m_1  m_2)^{5/7}} \ ,
\end{align}
where we have introduced $R_{\text{cl}}(z)$, a dimensionful factor (in units of 
$\rm{Gpc}^{-3}\rm{yr}^{-1}$) containing clustering information, following~\cite{Carr:2023tpt}.  It includes the PBH halo mass function, the dependence on the halo mass and radius, as well as the PBH cluster evolution with redshift. Details of the calculation of this factor can be found in Appendix \ref{App: Clustering Factor} that extend the previous estimations of~\cite{Carr:2023tpt}. We illustrate the dependence of $R_{\text{cl}}$ on $\fpbh$ for different values of $n_{\rm s}$ for $z = 0$ (today) in Fig.~\ref{Fig:rcl_fpbh}.

\subsection{SGWB from unresolved PBH binaries}
The SGWB from PBH mergers follows the standard relationship to the comoving number density of compact binary sources (for e.g. astrophysical black holes, neutron stars).  We summarise this derivation here following Ref.~\cite{Phinney:2001di} (see also e.g. Refs.~\cite{Bagui:2021dqi, Clesse:2016ajp, Mandic:2016lcn, Franciolini:2022tfm} for a similar derivation applied to the PBH scenario).

The spectrum of GWB expected from  PBH binaries (PBHBs) today can be written as

\begin{align}\label{Eq: GW energy fraction}
     \Omega_{\text{gw},0}^{\rm PBHB}(f) = \frac{1}{\rho_c} \frac{{\rm d} \rho_{\rm gw}^0}{{\rm d} \log f} =\frac{1}{\rho_c} \int_0^{\infty}\, N(z) \frac{1}{1+z} \, \bigg(f_{\rm r} \frac{{\rm d} E_{\rm gw}}{{\rm d} \, f_{\rm r}} \bigg) \, {\rm d} z,
\end{align}

 \noindent where $N(z)$ is the comoving number density of PBH merger events between $z$ and $z+\rm d z$ and $f_{\rm r} \,( {\rm d} E_{\rm gw}/d f_{\rm r})$ is the energy emitted as gravitational radiation per event between the frequency $f_r$ and $f_r + {\rm d} f_r$, with $f_r = f(1+z)$, the frequency measured in the source frame and $f$ the observed GW frequency today\footnote{Note that the differential energy ${\rm d} E_{\rm gw}/{\rm d} f_{\rm r}$ emitted per event is expressed in the source rest frame as well. This explains the additional factor of $(1/1 + z)$ in Eq.~\eqref{Eq: GW energy fraction}, which accounts for the GW redshifting.  This follows from the scaling of GW energy density as $\rho_{\rm gw} \propto a^{-4}$.}. 

We can relate $N(z)$ to the total merger rate per comoving unit volume $\mathcal V(z)$ through 
\begin{equation}
N(z) = \frac{\mathcal V(z) }{ [H(z) (1+z)] }
\end{equation}
where 
$H(z) = H_0 \, ( \Omega_{\rm m}  ( 1 + z)^3 + \Omega_{\rm r} ( 1 + z)^4 + \Omega_{\Lambda} )^{1/2}$, with $\Omega_{\rm m} =  0.315$ and $\Omega_\Lambda = 0.685$.   

For the case of circular PBH binaries of masses $m_1$ and $m_2$, in the Newtonian limit, the form of the differential GW spectrum emitted in Eq.~\eqref{Eq: GW energy fraction} can be well-approximated by

\begin{equation}\label{eq:gw_inspiral}
      \frac{{\rm d} E_{\rm gw}}{{\rm d}  f_{\rm r}} = \frac{\pi^{2/3}}{3 G} \frac{(G \, \mathcal{M}_c)^{5/3}}{f_{\rm r}^{1/3}} \, ,
\end{equation}
where $\mathcal{M}_c$ is the chirp mass defined in terms of component PBH masses as $\mathcal{M}_c^{5/3} = m_1 m_2(m_1+m_2)^{-1/3}$.  Eq.~\ref{eq:gw_inspiral} is valid for $f_{\rm r} < 2 f_{\rm ISCO}$  where 
\begin{equation}
f_{\rm ISCO} = 2.3 \, {\rm {k Hz}} \, \bigg(\frac{M_{\odot}}{ m_1+m_2 } \bigg)
\end{equation}
is the frequency corresponding to the Innermost Stable Circular Orbit (ISCO).  In principle, there is a lower limit on frequency  coming from the initial separation of PBH binaries, but for the typical velocity dispersions in the PBH clusters we consider, this lower limit is below the nHz range~\cite{Clesse:2016ajp}. 

Note that the GW spectrum given above captures the inspiral phase of any BH binary (including PBH binaries), while for the merger and ringdown phases, a completely different dependence of the spectrum on $f_{\rm r}$ is expected (see for example expressions in Ref.~\cite{Zhu:2011bd}). For PTA frequencies of $\mathcal{O} \, ({\rm nHz})$, the leading contribution to the GW spectrum comes from inspiralling BHs.

We can finally express the GW spectrum from PBH binaries by plugging Eq.~\eqref{Eq:Merger Rate} and Eq.~\eqref{eq:early_merger} in Eq.~\eqref{Eq: GW energy fraction}, giving

\begin{align}\label{eq:omega_gw_merg}
    \Omega_{\text{gw},0}^{\rm PBHB}(f) & =  f^{2/3} \, \frac{(\pi \, G)^{2/3}}{\rho_c} \, \int {\rm d} z  \, {\rm d} \ln m_1 {\rm d} \ln m_2 \\ \nonumber
    & \times \frac{1}{H(z) (1 + z)^{4/3}} \frac{{\rm d} \, \mathcal{V}(z)}{{\rm d} \ln m_1 {\rm d} \ln m_2} \mathcal{M}_c^{5/3}
\end{align}
with $\mathcal{V} = \mathcal{V}_{\rm early} + \mathcal{V}_{\rm late}$.
At frequencies lower than $2 f_{\rm ISCO}$, the GW spectrum has the characteristic frequency dependence, $\Omega_{\text{gw}}(f) \sim f^{2/3}$, as expected for inspiralling binaries under the approximation of circular orbits~\cite{Phinney:2001di}.  


\section{Results and interpretation \\of PTA data}\label{Sec: PTAs}

In this section, we describe our analysis of the PTA dataset for a GW background sourced by a combination of scalar perturbations and PBH binaries. 
For fixed values of the primordial parameters $(A_{\zeta}, n_{\rm s} , k_{\rm min}, k_{\rm max})$ introduced in Eq.~\eqref{Eq: Primoridal PS}, the GWB is given by summing Eq.~\eqref{Eq: Analytical Omega_gw} and Eq.~\eqref{eq:omega_gw_merg}, i.e., 
\begin{equation}\label{eq:omega_total}
h^2 \Omega_{\rm gw}(f) = h^2 \Omega_{\rm gw}^{\rm SIGW}(f) +  h^2\Omega_{\rm gw}^{\rm PBHB}(f).
\end{equation}
In a general PBH model, our search emphasizes the contribution to the GWB from SIGWs in addition to the GWB from PBH binaries. However, there can be alternative models with a suppressed SIGW background or PBH formation scenarios which do not lead to a GWB. In such cases, the total GWB would originate only from PBH binaries. We conduct a separate analysis to cover this scenario.

We perform Bayesian analyses by obtaining Monte Carlo Markov Chains (MCMC) using the publicly available code {\tt PTArcade} \cite{Mitridate:2023oar, andreamitridate_2023} in which we implemented the GW signal $h^2 \Omega_{\rm gw}(f)$ in Eq.~\eqref{eq:omega_total} from PBH mergers and scalar induced GW. In our search, we used the latest IPTA dataset (IPTA DR2) \cite{Perera:2019sca} which contains timing data from 65 pulsars including a combination of the following individual PTA data releases: the EPTA data release 1.0~\cite{EPTA:2016ndq}, the NANOGrav 9-year data set~\cite{2015}, and the PPTA first data release~\cite{Manchester_2013, Reardon:2015kba}. We furthermore include the pulsar white and red noise contributions according to the prescription of IPTA DR2~\cite{Antoniadis:2022pcn}. Following the prescription in Ref.~\cite{Antoniadis:2022pcn}, we limit the search of GW background to the first 13 frequency bins. 

Our analysis is divided into two parts.  For the analysis in Section~\ref{SubSec: Fixed Rcl}, we compute the clustering factor $R_{\rm cl}$ appearing in the late-time PBH merger rate in Eq.~\eqref{Eq:Merger Rate} as a function of redshift $z$ from the primordial power spectrum parameters, following Appendix ~\ref{App: Clustering Factor}, such that at any $z$, $R_{\rm cl} \equiv R_{\rm cl}(A_{\zeta} , n_s , k_{\rm min}, k_{\rm max})$ is fixed.  While for the analysis in Section~\ref{SubSec: Free Rcl}, we consider the clustering factor $R_{\rm cl}$ appearing in Eq.~\eqref{Eq:Merger Rate} as an extra free parameter in addition to the primordial spectrum parameters, such that the total GWB is now given by $\Omega_{\rm gw}(f; R_{\rm cl}, A_{\zeta} , n_s , k_{\rm min}, k_{\rm max})$.  This allows us to consider models with enhanced PBH clustering.  Here, we also include a scenario where the entire GW background arises from PBH binaries only.

\subsection{Fixed Clustering Factor Analysis}\label{SubSec: Fixed Rcl}

\begin{figure*}[t!]
     \centering
     \includegraphics[width=8.5cm]{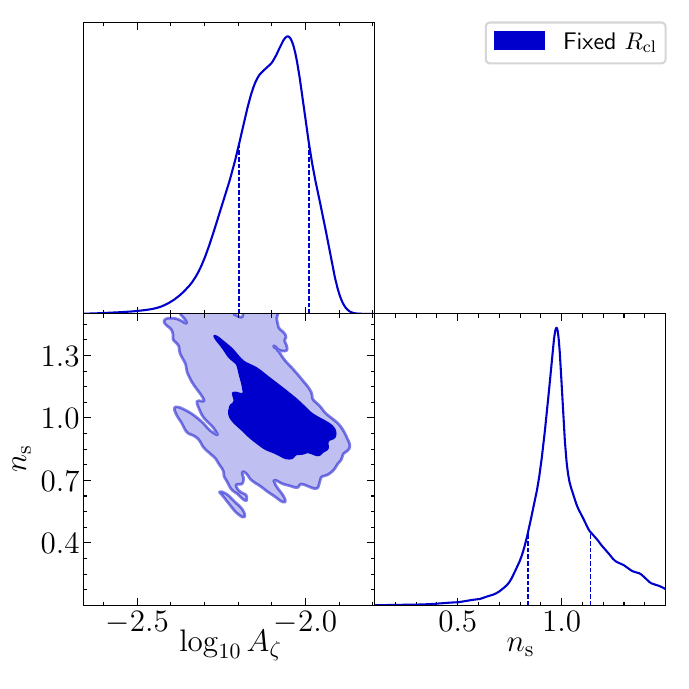}
     \includegraphics[width=8.5cm]{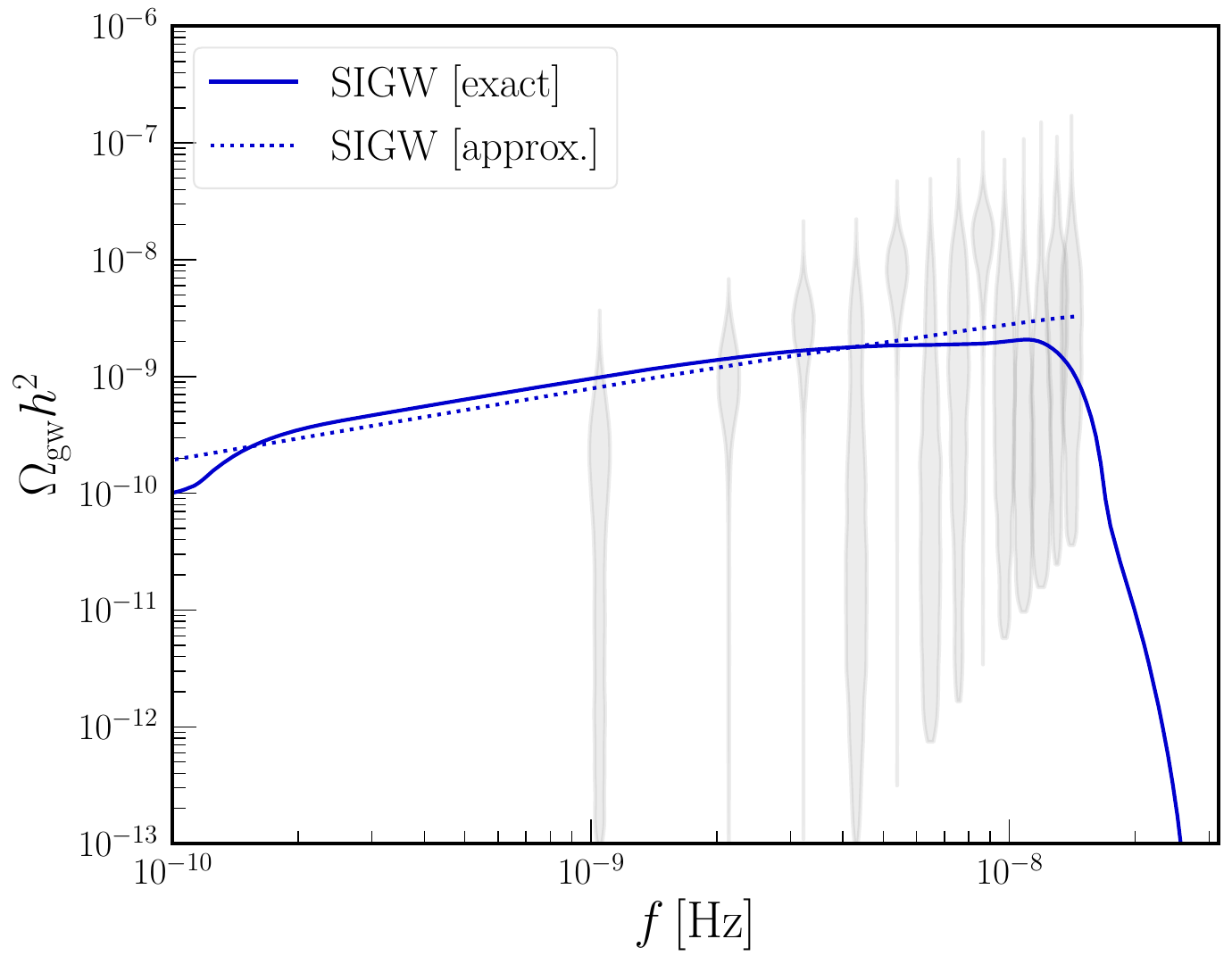}
    \caption{\textit{Left:} Posterior distributions for the parameters $A_{\zeta}$ and $n_{\rm s}$ of the primordial power spectrum defined in Eq.~\eqref{Eq: Primoridal PS} for the fixed clustering analysis (blue). The 1d marginalized distributions are reported on the diagonal of the corner plot, with the 68 \% Bayesian credible intervals (vertical lines), 
   while the off-diagonal panel (2d distribution) shows the 68 \% (darker) and 95 \% (lighter) C.I. regions respectively. The posteriors shown respect $\fpbh \leq 1$ as well as the CMB $\mu$ distortion limits from FIRAS/COBE~\cite{Mather:1993ij, Fixsen:1996nj}. \textit{Right:} GW spectrum obtained for the maximum likelihood values of parameters ($\log_{10} A_{\zeta} = -2.11, n_{\rm s} = 1.31, \log_{10} k_{\rm min} = 4.80,  \log_{10} k_{\rm max} = 6.96$) for the fixed $R_{\rm cl}$ analysis, containing contribution only from scalar induced GWs (see main text). Gray violins shown indicate the free spectrum posteriors obtained by converting the results of IPTA DR2~\cite{Antoniadis:2022pcn}.}
    \label{Fig: Posteriors Fixed _Rcl}
\end{figure*}

In this first analysis, the clustering factor $R_{\rm cl}$ multiplying the late PBH merger rate in Eq.~\eqref{Eq:Merger Rate}, is computed using the PBH halo mass function including Poissonian fluctuations (Eq.~\eqref{eq:rcl_defn}) for fixed primordial power spectrum parameters. The total GW spectrum in Eq.~\eqref{eq:omega_total} is only a function of the primordial parameters, $\Omega_{\rm gw} \equiv \Omega_{\rm gw}(f; A_{\zeta}, n_{\rm s}, k_{\rm min}, k_{\rm max})$.

We derive the posterior distributions of the primordial parameters $A_{\zeta}, n_s , k_{\rm min}, k_{\rm max}$ after setting logarithmic uniform priors for $A_{\zeta}, \ k_{\rm min}, \ k_{\rm max}$ and uniform prior for the spectral index $n_{\rm s}$, with prior limits given in Table~\ref{Tab: Prior Distribution}.
We furthermore restrict our analysis to the parameter space constrained by $f_{\rm PBH} \leq 1$ and where the primordial power spectrum respects the known constraint from the CMB $\mu$-distortion~\cite{Chluba:2012we} (see Appendix.~\ref{App: Constraints PS}).\\

The one- and two-dimensional posterior distributions for the fixed $R_{\rm cl}$ analysis are shown in blue in Fig.~\ref{Fig: Posteriors Fixed _Rcl}, where the light and dark regions represent 95 \% and  68 \%  C.I. regions respectively. The maximum posterior values and the 68 \% C.I. for the curvature amplitude are given by $\log_{10} A_{\zeta} = -2.06 \in [-2.20, -1.99],$ while for the spectral index are given by $\ n_{\rm s} = 0.97 \in [0.83, 1.14]$. From the full posterior distributions  shown in Fig.~\ref{Fig:fullpos_fixed}, we can see the maximum posterior values for the cut-off scales in the primordial spectrum:
$\log_{10} k_{\rm min}/\mathrm{Mpc}^{-1} = 5.68 \in [4.62, -]$ (where the upper $68 \%$ limit does not exist),  $\log_{10} k_{\rm max}/\mathrm{Mpc}^{-1} = 7.33 \in [6.73, \ 12.42]$.

For these maximum posterior values (see also Fig.~\ref{Fig: Posteriors Fixed _Rcl}), the PTA signal is always dominated by the scalar induced GWB, $\Omega_{\rm gw}^{\rm SIGW}(f)$ with the PBH binaries giving a much sub-leading contribution to the total GW spectrum. To explain the PTA common-spectrum process, a spectral index \( n_{\rm s} \sim 1.3 \) would yield the required increasing frequency dependence of the GW spectrum from SIGWs, \( \Omega_{\rm gw}^{\rm SIGW} \propto f^{2(n_{\rm s} - 1)} \). With the amplitude fixed at \( A_{\zeta} \sim 10^{-2} \) and \( k_{\rm max} \sim 10^7 \ \mathrm{Mpc}^{-1} \), a resulting signal of \( \Omega_{\rm gw} \sim \mathcal{O}(10^{-9}) \), covering the highest PTA frequency bin \( f_{\rm max} \sim 10^{-8} \ \mathrm{Hz} \) can be obtained.
 However, for these parameter values, light PBHs will overclose the Universe giving $\fpbh \geq 1$. This explains the preferred central values we obtain for the spectral index $n_{\rm s} \sim 0.97$. This $n_{\rm s}$ value results in a broad PBH mass distribution, where $k_{\rm max}$ is not restricted by the $\fpbh \leq 1$ constraint. This can be further seen in the 2d distribution for $(n_{\rm s}, k_{\rm max})$ in Fig.~\ref{Fig:fullpos_fixed}, where for $n_{\rm s} \sim 1,$ all $k_{\rm max} (k_{\rm min})$ values are allowed. Additionally, the CMB $\mu$-distortion limit becomes important for $\log_{10} k_{\rm min} \lesssim 4.4$, which explains the cut-off in the corresponding 2d distribution for $(n_{\rm s}, \log_{10} \ k_{\rm min})$.  Finally, the upper limit
on the curvature perturbation amplitude of $\log_{10} A_{\zeta} \leq -1.9$ is set by the $\fpbh \leq 1$ constraint. 
The resulting maximum likelihood GW spectrum for this analysis is shown Fig.~\ref{Fig: Posteriors Fixed _Rcl} (right) as solid (dotted) curves for the exact (approximate) GWB arising from the dominant scalar induced GW component obtained using Eq.~\eqref{eq:omegaGW_0} (Eq.~\eqref{Eq: Analytical Omega_gw}).

The PBH binaries in this analysis give a sub-leading contribution to the total GW spectrum as already mentioned. This can be understood as follows: the typical value of the clustering factor, \(R_{\text{cl}}\), from Eq.~\eqref{eq:rcl_defn}, is approximately $\mathcal{O}(1 - 100) \, \text{Gpc}^{-3} \, \text{yr}^{-1}\) for mass distributions with \(n_s \lesssim 1\) (see also Fig.~\ref{Fig:rcl_fpbh}). In contrast, for mass spectra with \(n_s > 1\), the clustering parameter can be large \(R_{\text{cl}} \sim 10^6 \, \text{Gpc}^{-3} \, \text{yr}^{-1}\) (see also Fig.~\ref{Fig:rcl_fpbh}), however, for these values of \(n_s\), the average PBH mass is typically \(\ll 1 M_\odot\) for \(k_{\text{max}} > 10^7 \, \text{Mpc}^{-1}\), leading to an overall small merger rate from late PBH binaries. Moreover, the CMB \(\mu\)-distortion limits set \(k_{\text{min}} \gtrsim 10^{4.4} \, \text{Mpc}^{-1}\) which corresponds to a cut-off in the PBH mass function giving \(m \lesssim 1000 \ M_\odot\), resulting in a significant suppression of the late-time PBH merger rate in Eq.~\eqref{Eq:Merger Rate}. Consequently, the GW spectrum from PBH binaries, \(\Omega^{\text{PBHB}}_{\text{gw}}(f)\), is suppressed at nHz frequencies.

\begin{table}[b!]
\centering
\begin{tabular}{ l  c }
\hline
\hline \\[0.01cm]
Search &   Bayes Factors \\[0.03cm]
 \hline\\[0.03cm]
 SIGWs + PBHBs [Fixed $R_{\rm cl}$]   &  \quad $\log_{\rm 10} (B_{\rm SMBHB, PBH} )= 1.88$  \\[0.2cm]
 
 SIGWs + PBHBs [Free $R_{\rm cl}$]   & \quad $\log_{\rm 10} (B_{\rm SMBHB, PBH} )= 1.86$ \\ [0.2cm]

 PBHBs [Free $R_{\rm cl}$] & \quad  $\log_{\rm 10} (B_{\rm SMBHB, PBH} )= 0.05$  \\[0.2cm]
 
\hline
\end{tabular}
\caption{Summary table for Bayes factors $B_{\rm SMBHB, PBH}$ for the astrophysical SMBHB model with respect to the PBH model for the free and fixed clustering factor analyses.}
\label{Tab: Bayes Factors Fixed and Free}
\end{table}

Finally, we compare this PBH model with a fixed clustering factor to the astrophysical model of inspiralling SMBHBs. The GW background for the latter is characterized by a power-law given by
\begin{equation}
\Omega_{\rm gw}^{\rm SMBHB} = \frac{2 \pi^2}{3 H_0^2} A_{\rm BHB}^2 \bigg(\frac{f}{\rm{yr}^{-1}}\bigg)^{5 - \gamma_{\rm BHB}} \ \mathrm{yr}^{-2},
\end{equation}
where we have fixed $\gamma_{\rm BHB} = 13/3$~\cite{Phinney:2001di} and varied the amplitude $A_{\rm BHB}$, with log uniform priors for the latter given by $ -18 \leq \log_{10} A_{\rm BHB} \leq - 11$. 

We compare the two models using the Bayes factor, $\log_{10} B_{a,b}$, of model $a$ with respect to model $b$. With the IPTA DR2 analysis for fixed $R_{\rm cl}$ (see Fig.~\ref{Fig: Posteriors Fixed _Rcl}), we find: $\log_{\rm 10} (B_{\rm SMBHB,\rm PBH} ) \simeq 2$ (see Table~\ref{Tab: Bayes Factors Fixed and Free}), which implies very strong evidence for the astrophysical SMBHB model over the PBH merger model with fixed clustering parameter $R_{\rm cl}$ including SIGWs triggered at PBH formation.

\begin{figure*}[t!]
    \centering
    \includegraphics[width=8.5cm]
    {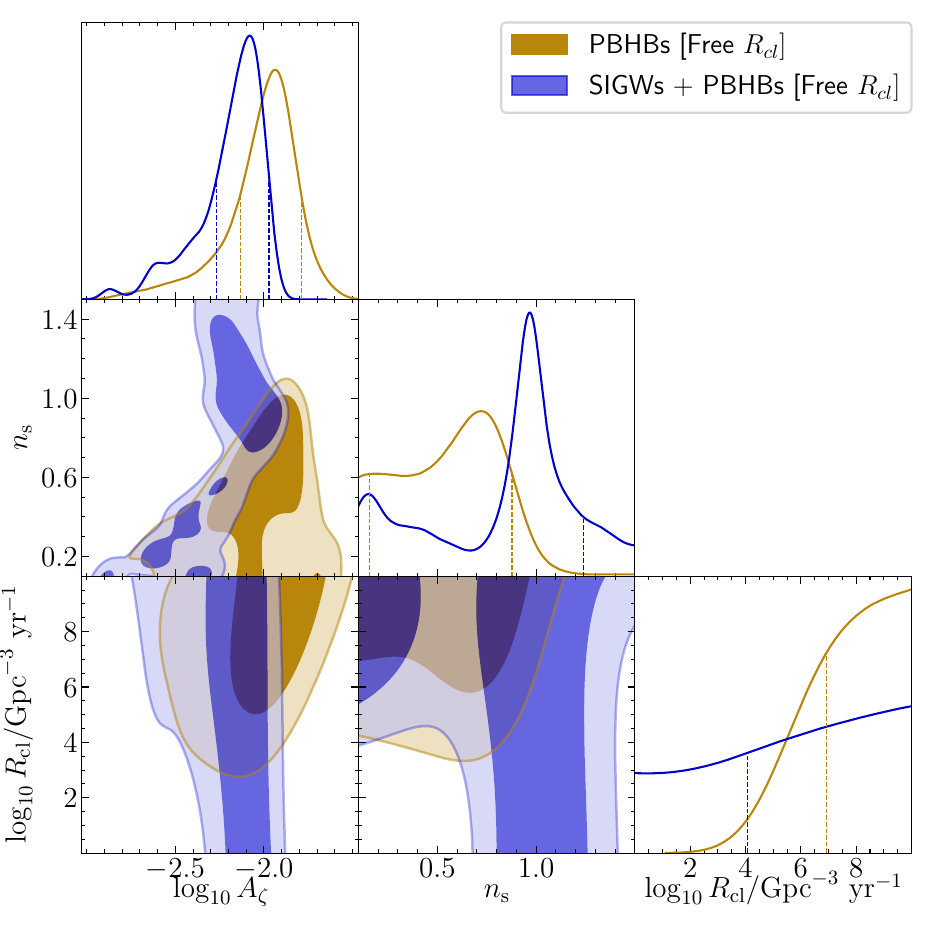}
    \includegraphics[width=8.5cm]{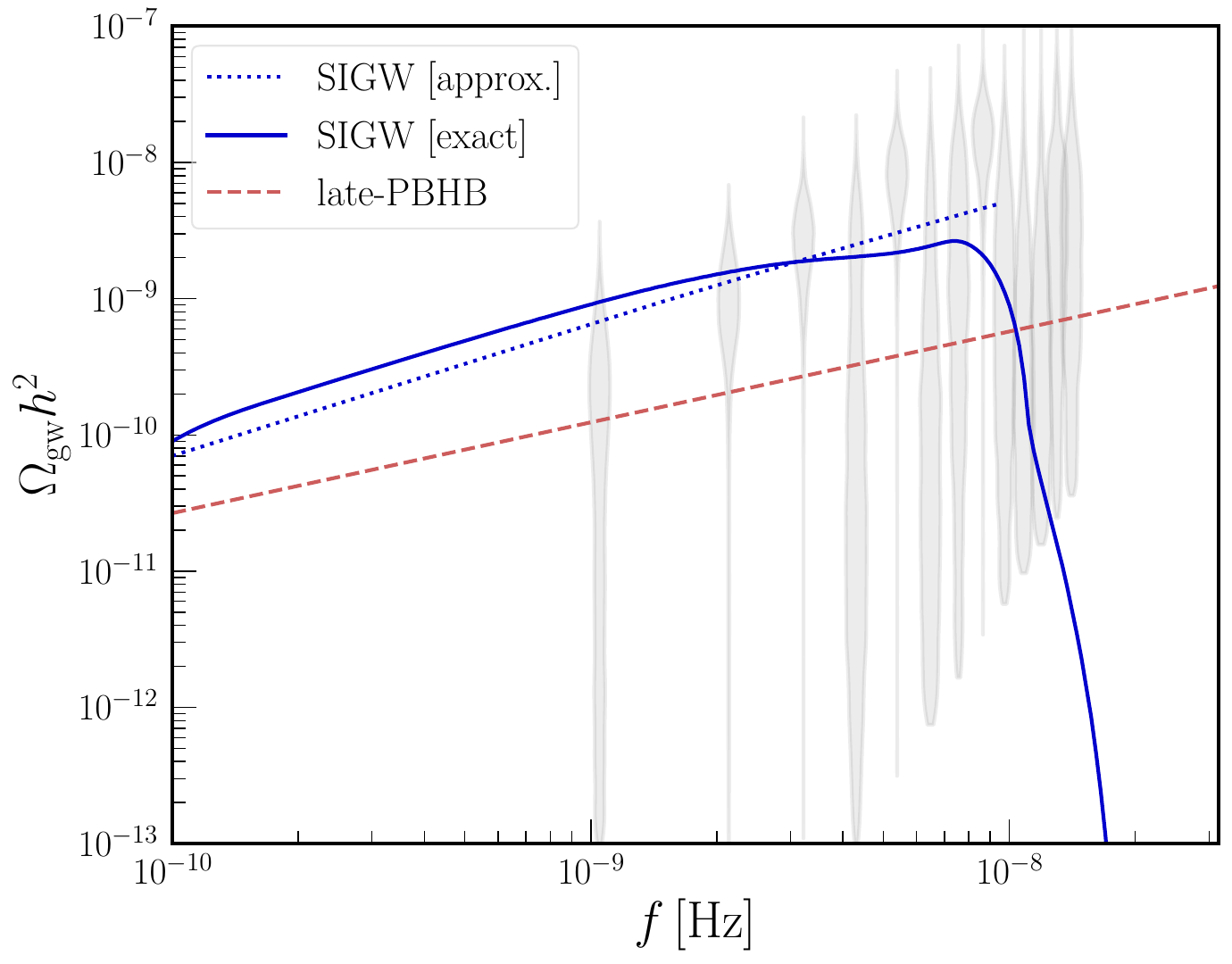}
    \caption{\textit{Left}: Posterior distributions for the parameters of the primordial power spectrum defined in Eq.\eqref{Eq: Primoridal PS} for the free clustering parameter $R_{\rm cl}$ analysis. The 1d marginalized distributions are reported on the diagonal of the corner plot 
   while the off-diagonal panel (2d distribution) shows the 68 \% (darker) and 95 \% (lighter) C.I. regions respectively. The blue and yellow contours correspond respectively to the analyses performed with SIGWs + PBH binaries and with PBH binaries only. \textit{Right}: Maximum likelihood GW spectrum for the free clustering analysis ($\log_{10} A_{\zeta} = -2.11, n_{\rm s} = 1.49, \log_{10} k_{\rm min} = 4.61,  \log_{10} k_{\rm max} = 6.78$) with contribution from both SIGWs and late PBH binaries. The signal is dominated at small frequencies by the SIGWs (blue), scaling as $\Omega_{\rm gw}^{\rm SIGW} \sim f^{2(n_s-1)}$, and from $f > f_{\rm max} \sim 10^{-8} \ \mathrm{Hz}$ onwards, by GWB from inspiralling (late) PBH binaries (dashed red) with the characteristic scaling $\Omega_{\rm gw}^{\rm PBHB} \sim f^{2/3}$.}
    \label{Fig:free_pos_part}
\end{figure*}

\subsection{Free Clustering Factor Analysis}\label{SubSec: Free Rcl}

In this section, we describe the analysis in which the clustering factor $R_{\rm cl}$ in the PBH merger rate (in PBH clusters) in Eq.~\eqref{Eq:Merger Rate} is considered to be a free parameter.   The total GW spectrum in Eq.~\eqref{eq:omega_total} also depends on the clustering parameter now i.e., $\Omega_{\rm gw} \equiv \Omega_{\rm gw}(f; A_{\zeta}, n_{\rm s}, k_{\rm min}, k_{\rm max}, R_{\rm cl})$. This can help determine the enhancement in late-time PBH merger rates required to fit the PTA signal, and verify whether such large PBH merger rates can be realistic and accounted for with current uncertainties on the PBH clustering model (described in appendix~\ref{App: Clustering Factor}).

We consider a uniform logarithmic prior on the free clustering parameter: $\log_{10} R_{\rm cl} \in [0,10] $ (see also Table~\ref{Tab: Prior Distribution}). The $\mu$-distortion limits from COBE/FIRAS~\cite{Chluba:2012we} and $\fpbh \leq 1$ constraint are implemented as before.

The one- and two-dimensional posterior distributions for the free $R_{\rm cl}$ analysis including both contributions from SIGWs and PBH binaries to the GWB are shown in blue in Fig.~\ref{Fig:free_pos_part} (full posteriors shown in Fig.~\ref{Fig:fullpos_free}). The maximum posterior values and the 68 \% C.I. for the curvature amplitude is given by $\log_{10} A_{\zeta} = -2.10 \in [-2.28, -1.97]$. For the spectral index, the maximum posterior value is given by $\ n_{\rm s} = 0.97$ and the 68 \% C.I. upper limit is  $n_{\rm s} < 1.22$ (no lower limit exists). The free clustering parameter reaches the maximum posterior value at $\log_{10} R_{\rm cl}/ \mathrm{Gpc}^{-3} \mathrm{yr}^{-1} = 10$, (equal to the upper limit of the prior value) with the 68 \% C.I. lower limit given by $\log_{10} R_{\rm cl}/ \mathrm{Gpc}^{-3} \mathrm{yr}^{-1} > 4.10$. From the full posterior distributions  shown in Fig.~\ref{Fig:fullpos_free} (in blue), the maximum posterior values and the 68 \% C.I. for the cut-off scales in the primordial spectrum are:
$\log_{10} k_{\rm min}/ \mathrm{Mpc}^{-1} = 4.58 \in [4.39, \ 5.91], \ \log_{10} k_{\rm max}/ \mathrm{Mpc}^{-1} = 6.99 \in [6.69, \ 13.89]$. For these maximum posterior values, the GW spectrum is dominated by the scalar induced GWs and the contribution from (late) PBH binaries to the GW spectrum at $f \sim 1 \ \mathrm{nHz}$, is very suppressed, $\Omega_{\rm gw}^{\rm PBHB} \sim 10^{-17}$.

The 2-d posterior distribution for $(A_{\zeta} , n_{\rm s})$ shows two distinct regions in Fig.~\ref{Fig:free_pos_part} (left). The region with larger values of spectral index ($n_{\rm s} \geq 0.8$) corresponds to the parameter space for which the PTA signal is dominated by GWB from SIGWs. This can be seen from the 2-d posterior distribution of $(R_{\rm cl},n_{\rm s})$ in (left) Fig.~\ref{Fig:free_pos_part} where for $n_{\rm s} \sim 1,$ all values of $R_{\rm cl}$ are allowed. 
Whereas, the second region with lower values of the spectral index ($n_s \leq 0.8$) corresponds to a PTA signal fitted by a GWB contribution from both scalar induced GWs and PBH binaries. The clustering factor required to enhance the contribution from late PBH binaries is $ R_{\rm cl} \geq 10^4 \ \mathrm{Gpc}^{-3} \ \mathrm{yr}^{-1}$, much larger than the expected value for Poisson fluctuation induced PBH clustering (see typical values in Fig.~\ref{Fig:rcl_fpbh}). Moreover, the late PBH binaries in this analysis can only contribute to the total GWB when $\rcl \fpbh^2 \sim \mathcal{O}(1)$. This can explain the large maximum posterior value of clustering factor, $R_{\rm cl} \sim 10^{10} \ \gpcy$  obtained, which compensates for the small $\fpbh$ values. Since $\Omega_{\rm gw}^{\rm PBHB} \propto \fpbh^2$ for late PBH binaries, changing $\log_{10} \ A_{\zeta}$ even slightly leads to a drastic change in the GW contribution from late binaries, whereas the scalar induced spectrum remains almost constant due to the quadratic dependence on the power spectrum amplitude $\osig \propto A_{\zeta}^2$.

The resulting maximum likelihood GW spectrum for this analysis is shown Fig.~\ref{Fig:free_pos_part} (right)  where the total GW background is a combination from scalar induced (shown in blue solid) till the cut-off $k_{\rm max} (f_{\rm max}) \sim 10^7 \ \mathrm{Mpc}^{-1} (10^{-8} \ \mathrm{Hz})$. For the higher frequency  bins $f > f_{\rm max}$, the PTA signal is fitted with the contribution from (late) PBH binaries (shown in dashed red).
A strong preference is found in favor of the astrophysical model with only inspiralling SMBHBs in comparison to this PBH model with SGWB from PBH binaries and associated SIGWs, with the Bayes factor $\log_{\rm 10} (B_{\rm SMBHB, PBH} ) = 1.86$. \\[0.3cm]


Finally, a similar analysis is performed with the free $R_{\rm cl}$ parameter including the contribution from PBH binaries (late and early) only (without the SIGWs contribution).  The 1d, 2d posterior distributions are shown in the left panel of Fig.~\ref{Fig:free_pos_part} in yellow. The maximum posterior values for all parameters and the corresponding 68 \% C.I. are: $\lnAz = -1.971 \in [-2.11, -1.78]$, $n_{\rm s} = 0.72 \in [0.16 , 0.88]$, $\lnrcl = 10$ (equal to the upper prior limit) with the lower 68 \% limit given by $\lnrcl \geq 6.87$. From the full posteriors shown in Fig.~\ref{Fig:fullpos_free} (in yellow), the maximum posterior values for the cut-off scales in the spectrum are:  $\lnkmn = 4.91 \in [4.54, \ 5.48]$ and  $\lnkmx = 6.48 \in [6.05, 17.83]$. Since in our model, the $\rcl$ parameter enters only in the late binary GWB, the only dominant contribution comes from the late PBH binary channel. The total PBH abundance $\fpbh$ for these maximum posterior values is 0.1 and for a large range of PBH masses, the posterior PBH abundance is strongly constrained from CMB (see Fig.~\ref{fig:f_m_constraints} in the next section and the related explanation).


\section{PBH Constraints}
\label{ref:pbh_constraints}

\begin{figure*}[t!]
    \centering
    \includegraphics[width=11cm]{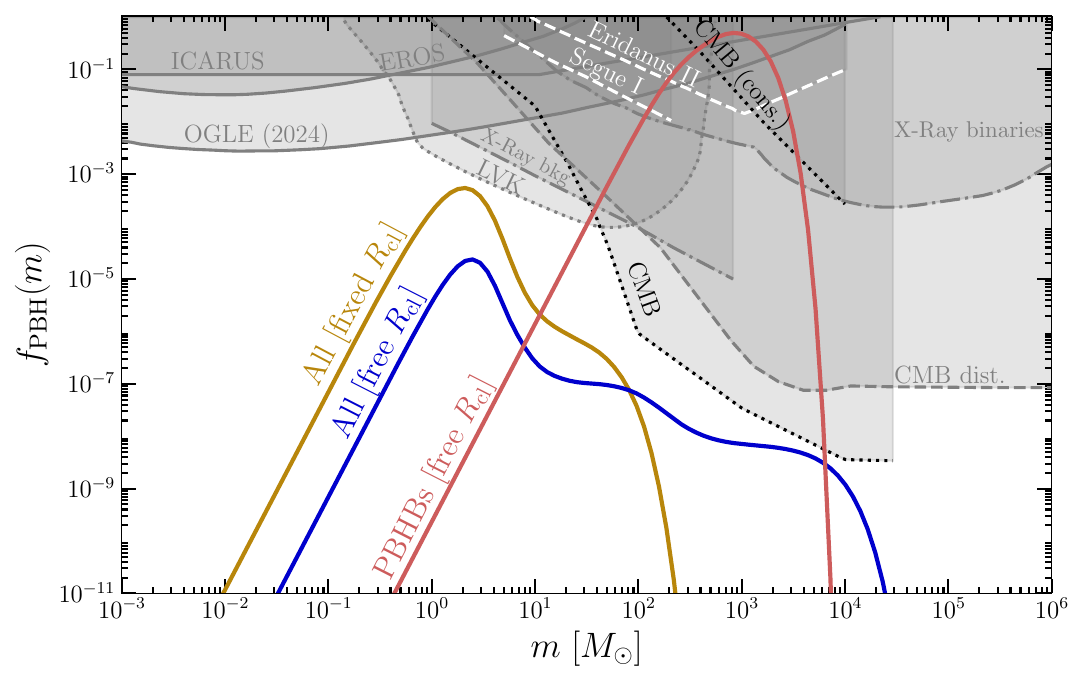}
    \caption{The PBH mass distribution derived using the maximum posterior values of our model parameters obtained from the PTA inference this work, compared to existing (most stringent) constraints on PBHs~\cite{LISACosmologyWorkingGroup:2023njw}. Among existing PBH constraints, we have shown microlensing ones from EROS~\cite{Macho:2000nvd}, ICARUS~\cite{Oguri:2017ock} and OGLE~\cite{Mroz:2024mse} (solid gray), dynamical limits from the heating of dwarf galaxies Eridanus II and Segue I (white dashed), constraints from X-ray binaries~\cite{Inoue:2017csr} and X-ray background (dot-dashed gray) [X-ray bkg] and finally constraints coming from CMB distortions [CMB dist.] (gray dashed) as well as the most stringent~\cite{Serpico:2020ehh} [CMB] and conservative constraints~\cite{Agius:2024ecw} [CMB (cons.)] from CMB anisotropies (dotted black). Constraints from Ligo-Virgo-KAGRA O3 merger rates (dotted gray) shown here have been taken from Ref.~\cite{Andres-Carcasona:2024wqk}.}
    \label{fig:f_m_constraints}
\end{figure*}

In this section, we describe the constraints on the PBH $(\fpbh(m), m)$ parameter space obtained from our analysis with IPTA DR2 in relation to existing PBH constraints. 
Fig.~\ref{fig:f_m_constraints} shows the posterior predictive distribution for the PBH abundance $\fpbh(m)$ (as defined in Eq.~\eqref{Eq: Mass function}) obtained from the PTA inference, alongside the most stringent existing constraints (see Ref.~\cite{Carr:2020gox} for a recent review on this) in this PBH mass range. \\
In the planetary-mass and low stellar-mass range up to $m \sim 10 \ M_{\odot}$, the strongest constraints on the PBH abundance $\fpbh(m)$ come from microlensing surveys. These include: EROS searches for massive compact halo objects (MACHOs) towards the Large Magellanic Cloud (LMC)~\cite{Macho:2000nvd}, fast transient event (ICARUS) near critical curves of massive clusters~\cite{Oguri:2017ock, Kawai:2024bni}, and the most recent results from observation of stars in LMC and Galactic Bulge by the Optical Gravitational Lensing Experiment (OGLE)~\cite{Mroz:2024mse}. In the stellar and intermediate mass ranges, the most stringent limits come from X-ray emissions from PBHs interacting with the interstellar medium. In Fig.~\ref{fig:f_m_constraints}, we show bounds coming from observation of X-ray binaries~\cite{Inoue:2017csr}, cosmic radio and cosmic X-ray backgrounds from ARCADE2 and Chandra~\cite{Ziparo:2022fnc}. More stringent limits from LVK black hole population analysis~\cite{Andres-Carcasona:2024wqk} exist but they do not apply to our mass distributions with $n_{\rm s} \gtrsim 0.9$.\\
In the intermediate mass range, the most severe constraints are coming from CMB anisotropies produced by accreting PBHs in the early Universe, where we are showing the bounds from Ref.~\cite{Serpico:2020ehh}. Note that in general bounds on PBHs from CMB anisotropies can change drastically on changing the PBH accretion model and on including DM mini-halos, and they additionally suffer from astrophysical uncertainties. Ref.~\cite{Agius:2024ecw} set a conservative limit on PBHs from CMB where for $2 M_{\odot} \lesssim m \lesssim 10^4 M_{\odot}$, the PBH abundance is constrained to be $10^{-3} \lesssim f_{\rm PBH} \lesssim 1$ (see also Ref.~\cite{Facchinetti:2022kbg}). Finally, we have shown Dynamical limits coming from the distribution of stars in Segue I~\cite{Koushiappas:2017chw} and Eriadnus II~\cite{Brandt:2016aco} dwarf galaxies in Fig.~\ref{fig:f_m_constraints}. 

The predictive PBH abundance distributions shown in Fig.~\ref{fig:f_m_constraints} correspond to the PBH mass distributions for the maximum posterior parameter values obtained from the Bayesian inference of the IPTA DR2. 
In Fig.~\ref{fig:f_m_constraints} the solid red curve corresponds to the scenario where only late-time PBH binaries contribute to the entire GWB and the clustering parameter, $\rcl$, is free.  The dark blue (yellow) curves correspond to the scenario where all channels, SIGWs, early PBHBs and late PBHBs, are included in the GWB analysis for free (fixed) clustering parameter $R_{\rm cl}$, respectively.  We observe that the distribution obtained for only PBH binaries with free $R_{\rm cl}$ is excluded by several probes, including the most conservative CMB limits. Moreover, the posterior values for this analysis give $\fpbh \sim 0.8, \ \langle m \rangle \sim 600 \ M_{\odot}$, and are in violation of the constraints coming from LVK merger rates $\sim \mathcal{O}(1 - 100) \ \mathrm{Gpc}^{-3} \mathrm{yr}^{-1}$~\cite{LIGOScientific:2017bnn}. \\
The two other analyses including SIGWs, however, evade all these limits and are viable.  The posterior values for the fixed and free $R_{\rm cl}$ analysis, including the contribution from both SIGWs and PBH binaries to the GWB, peak at $f_{\rm PBH} \sim 10^{-5} - 10^{-4}$, at PBH mass of around $4 - 10 \ M_\odot$.  Moreover, the merger rates of both early and late PBH binaries in these two scenarios are compatible with the LVK constraints for merger rates (see also Ref.~\cite{Andres-Carcasona:2024wqk}). This can open some interesting perspectives for PBH models to explain at the same time a small fraction of LVK mergers and the GWB observed by PTAs.


\section{Conclusion and Discussion}\label{Sec: Conclusion}

In this work, we have searched for a GW background arising from inspiralling PBH binaries in the International PTA dataset (IPTA DR2). We have considered two channels for PBH binary formation in our analysis: PBH pair formation in the early Universe as well as PBH binary formation via dynamical capture in late-Universe PBH clusters. While for small PBH abundance or $\fpbh$, PBH binaries decoupled in the early Universe from the Hubble flow can dominate the merger rate, for large values of $\fpbh$, PBH binaries formed dynamically inside PBH clusters can be dominant. Our Bayesian search in the IPTA DR2 was largely motivated by the results of Ref.~\cite{Bagui:2021dqi} which showed that a GWB from late-time PBH binaries could be large enough to be probed by PTAs at the nHz frequency range, due to a large contribution from binaries with asymmetric masses (between a solar-mass PBH and and an intermediate-mass PBH). Our analysis shows that such a contribution to the GWB is largely suppressed due to the need to introduce a cut-off at scale $k_{\rm min} \sim 10^5 \ \mathrm{Mpc}^{-1}$ to evade the limits from CMB $\mu-$distortions~\cite{Chluba:2012we}, which restricts the formation of PBH with mass $m \gtrsim 1000 \ M_{\odot}$. 

In comparison to previous works, our analysis is based on merger rates for a broad PBH mass function including the features from the QCD epoch (as detailed in sec.~\ref{Sec: Mass function}). The broad PBH mass function used in our analysis has been derived assuming the direct collapse of enhanced density perturbations obtained from a power-law power spectrum of curvature perturbations at small scales, with a large-mass
cut-off (or small k, $k_{\rm min}$) introduced to evade the constraints from CMB $\mu$-distortions and a low-mass cut-off (or large k, $k_{\rm max}$) to prevent overclosing the Universe with light PBHs. 
Most importantly in relation to previous works, in our analysis for the search for GWB from PBH binaries, we have included the scalar induced stochastic GW background, which will be unavoidable due to the enhanced curvature perturbations needed for PBH formation in the standard scenario. 

We also note some limitations in our computation for merger rates both from early binaries and late binaries. For broad PBH mass spectra obtained for $n_{\rm s} \sim 0.96$, the early merger rate suppression from third-PBHs is still uncertain~\cite{Bagui:2021dqi}. 
For the late PBH binaries, we have carefully accounted for the formation of PBH binaries in late small scale halos with the correct halo mass function for Poissonian fluctuations, and included effects from dynamical heating~\cite{Brandt:2016aco, Carr:2023tpt} on the cluster size. Both these effects, as we also discuss in Appendix~\ref{App: Clustering Factor}, can have an important impact on the late merger rates. This latter effect of dynamical heating, in addition to, effects of mass segregation between light PBHs and heavy PBHs can lead to uncertainties on the late merger rate computation for broad PBH mass distribution (obtained for $n_{\rm s} \sim 1$). Our work, nevertheless, provides a rigorous and complete template for merger rate computation that can be used for analysis involving PBH merger rates, for instance for analysing limits on the GW background by LIGO-Virgo-Kagra.

Our main results from the Bayesian inference show that the contribution to the GWB from PBH binaries (both early and late channels) is largely subdominant with respect to the GWB from scalar induced perturbations for power law PBH mass distributions (see Sec.~\ref{SubSec: Fixed Rcl}). It is not possible to explain the PTA signal with only PBH binaries in late time PBH clusters, under the assumption of standard PBH formation scenario with PBH clustering from Poissonian fluctuations. \\
We further performed an analysis with a nuisance parameter in the PBH merger rate from late binaries given by the ``free'' clustering parameter $R_{\rm cl}$ (see Sec.~\ref{SubSec: Free Rcl}). We find that the values of $R_{\rm cl}$ required to explain the PTA signal with late PBH binaries is $\sim \mathcal{O}(10^4) \ \mathrm{Gpc}^{-3} \ \mathrm{yr}^{-1}$ larger than the $\rcl$ values arising from Poissonian fluctuations. Nevertheless, this analysis with the free clustering parameter $R_{\rm cl}$ can serve as a proxy for the enhancement in PBH clustering required for the GWB from late PBH binaries to dominate over the SIGW background. An enhancement in PBH clustering will necessarily impact the early PBH binary merger rates as well (see Refs.~\cite{PhysRevD.99.063532, Raidal:2017mfl, Ballesteros:2018swv}). However, our analysis, which uses a free clustering parameter, neglects this effect in the early PBH merger rate as it is difficult to estimate. \\
Our analysis also reveals that PBH mass distributions with $n_{\rm s} \lesssim 1$ are statistically preferred, because they naturally suppress low-mass PBHs (from overclosing the Universe) and do not require a small-scale cut-off $k_{\rm max}$.

In addition to the two channels considered in our analysis, a third channel via three-body interaction can also lead to PBH binary assembly in small scale PBH halos. Ref.~\cite{Franciolini:2022ewd} has shown that this can give a larger contribution than the  capture channel in small clusters. Given the required enhancement of merger rates to match the PTA signal, it is unlikely that the inclusion of this channel will have an impact on our main conclusions. Additionally, we always consider non-eccentric or circular binaries. The dynamical channel for PBH binary formation can lead to highly eccentric binaries as shown in Ref.~\cite{Cholis:2016kqi}. A more detailed analysis comparing all channels and incorporating eccentric binaries is left for a future work. 

In some sense, our analysis in this work contributes to the search for the cosmological origin of the PTA signal by excluding a series of plausible scenarios related to PBHs. As a corollary of our analysis, we can identify the following scenarios that can possibly lead to an increase in the merger rates from PBH binaries in the late-Universe: an increased clustering at PBH formation could lead to an increased merger rates from both early PBH binaries as well as late-time dynamically formed PBH binaries. In addition, an effect opposing the dynamical heating leading to contraction of the PBH cluster radius (such as gravitational cooling) could also enhance the merger rate. Note that such such enhanced PBH clustering could also have a significant effect on the microlensing constraints on PBH abundance (see for example Ref.~\cite{Gorton:2022fyb}). While for PBH formation from Gaussian perturbations, clustering at PBH formation is negligible, for non-Gaussian perturbations, an enhanced clustering can be expected as shown in Ref.~\cite{Franciolini:2018vbk}.\\
Moreover, in order to form PBHs with mass $m \gtrsim 1000 \ M_{\odot}$, which can eventually lead to increased merger rates and subsequently a larger GWB, a PBH formation mechanism is required that can evade the tight CMB spectral distortion limits on the primordial perturbation amplitude at scales larger than $k \sim 10^5 \ \mathrm{Mpc}^{-1}$. While in general, it is very challenging to form massive PBHs from direct collapse, these limits can be avoided, if the distribution of primordial curvature perturbations is highly non-Gaussian~\cite{Unal:2020mts}. Under this mechanism, PBHs can form from the smaller peaks in the primordial power spectrum from the tail of a sufficiently non-Gaussian distribution of the curvature perturbations $\zeta$. While, large scale observations on non-Gaussianity parameters $f_{\rm NL}$ and $g_{\rm NL}$ constrain $\zeta$ to be nearly Gaussian~\cite{Planck:2019kim}, small scales related to PBH formation are free from such constrains. See recent work in Ref.~\cite{Hooper:2023nnl} on the formation of supermassive PBHs from non-Gaussian $\zeta$. \\
An additional possible route to evade CMB spectral distortion limits that can additionally suppress the GWB from SIGWs is to consider another PBH formation mechanism such as PBHs produced from a first-order phase transition~\cite{Liu:2021svg}. 

In addition to $\mu$-distortion limits, our results are also strongly dependent on the condition that PBHs do not overclose the Universe, $\fpbh \leq 1$. Indeed, we have implemented this condition in obtaining our final posteriors shown in Sec.~\ref{Sec: PTAs} and App.~\ref{App: Posteriors}. Since the computation for PBH abundance is exponentially sensitive to the choice of the window function, the threshold for PBH formation (see also discussion in Ref.~\cite{Dandoy:2023jot}) and to the method of abundance computation, our posterior parameter values are subject to this theoretical uncertainty as well.

In the next few years, the combined dataset from all PTAs  will be released as International PTA Data Release 3 (IPTA DR3), the analysis of which will improve our understanding on the origin of the GW excess in the nHz frequency range. While the most plausible motivation for this excess comes from astrophysical SMBHBs, as has been shown in literature, the origin of this signal can arise also from many well-motivated new physics scenarios in the early Universe. Our work clarifies the premises and assumptions that would be needed to see a GW background from PBH mergers in the dataset.


\section*{Acknowledgments}
We would like to thank members of the IPTA and NANOGrav collaborations for useful comments on a preliminary version of this work. SV would also like to thank G. Franciolini, G. Domènech, S. Babak for useful discussions and comments. The work of SV is supported by the Excellence of Science (EoS) project No. 30820817 - be.h “The H boson gateway to physics beyond the Standard Model” and by the IISN convention 4.4503.15.  The work of S.C. is supported by an Incentive Grant for Scientific Research (MIS) from the Belgian Fund for Research FNRS, and by the IISN convention 4.4501.19. 

\appendix
\section{Scalar Induced GWs}
\label{app:sigw}
In this appendix, we provide the analytical expressions that are used to calculate the GW spectrum during the radiation era.  
It is obtained by solving the double integral 
%
\begin{equation}\label{Eq: General Omega_SIGW}
\begin{split}
    &\Omega_{\text{gw}, \, \text{RD}}(k,\eta)  = \frac{1}{12}\left(\frac{k}{a(\eta)H(\eta)}\right)^2 \int^{\infty}_0 {\rm d}t \int^{1}_{-1} {\rm d}s\,\\
    &\times \Bigg[ \frac{t(2+t)(s^2-1)}{(1-s+t)(1+s+t)} \Bigg]^2 I_{\rm RD}^2(u, v, k\eta) \,  \mathcal{P}_{\zeta}(u k ) \, \mathcal{P}_{\zeta}(v k),
\end{split}
\end{equation}
with $H(\eta)$,  $a(\eta)$ being the Hubble rate and the scale factor as a function of the conformal time $\eta$. An additional time dependence is contained inside the function $I_{\rm RD}(u, v, k\eta)$.  In the limit $k\eta \gg 1$, it is given by~\cite{Kohri:2018awv}

\begin{equation} \label{eq:I_rd_full}
    \begin{split}
     I_{\rm RD}(u,v,x \to \infty) &= \frac{1}{x} \left( \frac{3(u^2+v^2-3)}{4u^3v^3 }\right)\Big(\sin(x) \\ \Big( -4uv+ 
    &(u^2+v^2-3) \log \frac{|3  - (u+v)| }{|3  - (u-v)| } \Big) - \\
    &\pi (u^2+v^2 -3) \Theta (v+u -\sqrt{3})\cos(x)\Big).
    \end{split} 
\end{equation}
For the production of SIGWs from small scales primordial perturbations, it is reasonable to assume that all the relevant modes $k$ enter the horizon early in radiation domination. Hence, deep in radiation domination, those modes satisfy the condition $k\eta \gg 1$.   On taking the oscillation average after squaring Eq.~\eqref{eq:I_rd_full} above, the factors of $\sin^2(x)$ and $\cos^2(x)$ will simply give a factor of $1/2$. We can thus write the oscillation average as:
\begin{equation}\label{Eq: I averaged}
    \begin{split}
     \Bar{I}_{\rm RD}^2(u,v,x) &= \frac{1}{2x^2} \left( \frac{3(u^2+v^2-3)}{4u^3v^3 }\right)^2\Big( \Big( -4uv+ \\
    &(u^2+v^2-3) \log \frac{|3  - (u+v)^2| }{|3  - (u-v)^2| } \Big)^2 + \\
    &\pi^2 (u^2+v^2 -3) \Theta (v+u -\sqrt{3})\Big).
    \end{split}
\end{equation}
One can note that the time dependence in the pre-factor of Eq.~\eqref{Eq: General Omega_SIGW}, $k/(aH) = k\eta = x$,
cancels the time dependence of $\Bar{I}_{\rm RD}^2(u,v,x)$ encoded in $x^{-2}$ in Eq.~\eqref{Eq: I averaged}.
The GW spectrum in the radiation era is therefore time-independent, as expected since GWs scale as radiation $\rho_{\rm GW} \propto \rho_{\rm r} \propto a^{-4}$.\\


\section{Thermal average of binary cross-section}
\label{app:sigma_avg}
In this appendix, we give details on how to compute the thermal average of the binary-formation cross-section defined in Eq.~\eqref{Eq: cross section}.

As mentioned in the main text, following the velocity model of Ref.~\cite{Bird:2016dcv}, one can approximate the PBH relative velocity denoted by $v$ with a Maxwell-Boltzmann distribution

\begin{equation}\label{eq:VDF}
    P(v) = N \left[ \exp( - \frac{v^2}{v_{\rm esc}^2}) -  \exp( - \frac{v_{\rm vir}^2}{v_{\rm esc}^2}) \right] \ ,
\end{equation}
where the second term in the distribution introduces a cut-off at the virial velocity of the halo, defined as
\begin{equation} \label{eq:vvir}
    v_{\rm vir} \simeq \sqrt{\frac{G \mh}{r_{\rm h}}},
\end{equation}
while $N$ is a normalisation constant found by imposing $\int_0^{v_{\rm vir}} {\rm d}^3 v \, P(v) = 1$.
Here, $v_{\rm esc}$ is the PBH velocity dispersion in the halo, defined as the escape velocity at the radius $R_{\rm max} = C_{\rm m} R_{\rm s}$ with $C_{\rm m} = 2.1626$,
\begin{equation}
    v_{\rm esc} = \sqrt{\frac{G M(R < R_{\rm max})}{R_{\rm max}}} =v_{\rm vir}  \sqrt{ \frac{C}{C_{\rm m}} \frac{g(C_{\rm m})}{g(C)}}~,
\end{equation}
where $g(C_{\rm m}) = 0.46759$ and 
\begin{equation}
M(R < R_{\rm max}) = \int_0^{R_{\rm max}} 4 \pi r^2 \rho_{\rm NFW}(r) \, \rm{d} r ~.
\end{equation}
Using Eq.~\eqref{eq:VDF}, we can finally compute the thermal average of the cross-section as
\begin{align}
    \langle \sigma_{\rm bin} v \rangle & = 4 \pi \int_0^{v_{\rm vir}} \mathrm{d} \, v \, \sigma_{\rm bin} \, v^3 P(v) \\ \nonumber
    & = 
    2 \pi \bigg(\frac{85 \pi}{6 \sqrt{2}} \bigg)^{2/7} \frac{G^2 (m_1 + m_2)^{10/7} (m_1 m_2)^{2/7}}{c^{10/7}} \nonumber  \\   & \times v_{\rm vir}^{-11/7} \, \mathcal{F}\big[C_m, C(M_{\rm h}), g(C_m), g(C) \big]   
\end{align} 
where $\sigma_{\rm bin}$ is given by Eq.~\eqref{Eq: cross section} and where $\mathcal{F}[C_{\rm m}, C(M_{\rm c}), g(C_{\rm m}), g(C)] \equiv \mathcal{F}_{\rm avg}$ is a  complicated function of the parameters $C_{\rm m}$ and $ C$ that remains after the integration of the Maxwellian velocity distribution given in Eq.~\eqref{eq:VDF}.
To a very good approximation, $\langle \sigma_{\rm bin} v \rangle \approx 3 \, \sigma_{\rm bin}(v_{\rm vir}) \, v_{\rm vir}$.  Note that, picking any other velocity distribution could lead to a larger thermal average.  This uncertainty from the choice of the PBH velocity distribution will be captured by our free $\rcl$ clustering analysis (see section~\ref{SubSec: Free Rcl}).
 
\section{PBH Clustering} 
\label{app:pbh_clusters}
In this appendix, we discuss the Poisson fluctuations in the PBH number density that contribute as an additional term in the matter power spectrum, dominant at small scales.  As mentioned in the main text, within the Press-Schechter formalism, these fluctuations collapse and form PBH halos.  We first provide a calculation of the clustering factor introduced in Eq.~\eqref{Eq:Merger Rate}, resulting from this effect.  Then we discuss the dynamical heating of PBH clusters to obtain the minimum cluster mass $M_{\rm min}$.  

\subsection{Poisson fluctuations from PBHs}
Due to the discrete nature of PBHs, Poisson fluctuations are expected in their local number density~\cite{Meszaros:1975ef}.  These Poisson fluctuations give rise to isocurvature perturbations in the matter power spectrum that could be observable as a small-scale plateau~\cite{Afshordi:2003zb} (see also e.g. Refs.~\cite{Inman:2019wvr,Murgia:2019duy,Gong:2017sie,Kashlinsky:2016sdv,Carr:2023tpt} for analyses including Poisson fluctuation effects for PBH dark matter).  We have neglected other possible sources of isocurvature fluctuations and particle dark matter clustering.  

The cold dark matter power spectrum ${P}_{\rm CDM}(k)$\footnote{This is the dimensionful matter power spectrum written in terms of the usual dimensionless power spectrum as $\Delta^2 = k^3 P_{\rm CDM}/ 2 \pi^2$.} in our scenario is therefore the sum of contributions from the adiabatic perturbations and isocurvature perturbations originating from these Poisson fluctuations. The adiabatic perturbations include the standard contribution from the primordial power spectrum on large scales and the small-scale term introduced in Eq.~\ref{Eq: Primoridal PS}.
The additional isocurvature contribution expected from Poisson fluctuations, in the case of a monochromatic PBH distribution, is given by~\cite{Inman:2019wvr, Hutsi:2019hlw}, 
\begin{align}\label{eq:p_iso}
    {P}_{\rm iso}^{\rm PBH} &
   = \fpbh^2 n_{\text{PBH}}^{-1} D_{\rm PBH}^2(z) \\
    & \simeq 1.7 \times 10^{-3}  \fpbh \, \bigg(\frac{m}{3 \ \sm}\bigg) \  (1 + z)^{-2} \ \mathrm{Mpc}^{3}~, \nonumber
\end{align}
where $D_{\rm PBH}(z)$ is the growth factor of PBH induced perturbations that can be approximated analytically as~\cite{Inman:2019wvr}
\begin{equation}
    D_{\rm PBH}(z) = \bigg( 1 + \frac{3}{2} \ \frac{\Omega_{\rm DM}}{\Omega_{\rm m}} \ \frac{1 + z_{\rm eq}}{1 + z} \bigg)
\end{equation}
where $z_{\rm eq} \approx 3400$ is the redshift for matter-radiation equality, $n_{\text{PBH}}$ is the average number density of PBHs in a comoving volume and we have used $n_{\text{PBH}} \ m = \fpbh \Omega_{\rm DM}^0 \rho_c$ in the second line in Eq.~\eqref{eq:p_iso}, with $\Omega_{\rm DM}^0 = 0.265$, $\rho_{\rm c} = 2.77 \times 10^{11} h^2 \sm \mathrm{Mpc}^{-3}$ and $h=0.674$.
Note that the above equation holds only for $k > k_{\rm eq}$, which is fulfilled for realistic PBH masses. \\[0.1cm]

 In the case of our broad mass distribution, different PBH masses induce Poisson fluctuations and their combined effect requires the substitution of $f_{\rm PBH} m$ by its averaged value over the considered mass function or $\langle f_{\rm PBH} m \rangle$ in Eq.~\ref{eq:p_iso}~\cite{Carr:2023tpt}. This is given by
 \begin{equation}
     \langle f_{\rm PBH} m \rangle = \int m \fpbh(m) \ {\rm d} \ln m.
 \end{equation}
 
 The next step is to use the Press-Schechter formalism to study the mass and size of PBH clusters induced by these Poisson fluctuations.

\subsection{Cluster formation}

Here, we give an estimate of the size of the PBH clusters formed from the collapse of Poisson density perturbations.  
According to the Press-Schechter formalism, the density fluctuations associated to the matter power spectrum, including Eq.~\eqref{eq:p_iso}, decouple from the Hubble flow and collapse into proper clusters when the overdensity exceeds the critical threshold of $\delta_c \approx 1.686$. 

{We can write the mass of the  cluster contained within a sphere of radius $\lambda = 2 \pi / k$ as:
\begin{equation}\label{eq:mh_k}
    M_{\rm h}(\lambda) = \frac{4 \pi}{3} \lambda^3 \rho_{\rm m}^0 (1 + \delta_c) \simeq 2 \times 10^{11} \ M_{\odot} \bigg(\frac{\lambda}{\mathrm{Mpc}} \bigg)^3 (1 + \delta_c)
\end{equation}
where $\rho_{\rm m}^0$ is the matter density today.}  Then a cluster of mass $M_{\rm h}$, associated to modes for which the Poisson term dominates, will be expected to form around a redshift $z_{\rm h}$, given by \cite{Clesse:2020ghq}
%
\begin{align}\label{eq:zh}
    (1+z_{\rm h}) &\simeq 3 \times 10^{-3} \bigg( \frac{k}{\mathrm{Mpc}^{-1}} \bigg)^{3/2} \left(\frac{\fpbh \ m}{\sm}\right)^{1/2} \\ \nonumber & \simeq 30  \ \left( \frac{10^6 \sm}{M_{\rm h}} \frac{\langle m f_{\rm PBH}\rangle}{\sm}\right)^{1/2} .
\end{align}
%
{To estimate the redshift of cluster formation above, we have used the expectation that at $z_{\rm h}$,  $\Delta_{\rm PBH}^2 \equiv k^3 P_{\rm iso}^{\rm PBH}/2 \pi^2 \simeq \delta_c^2$ and used $P_{\rm iso}^{\rm PBH}$ defined in Eq.~\eqref{eq:p_iso} and further used the cluster mass defined in Eq~\eqref{eq:mh_k} associated with $k$ in the second line.}

\subsubsection{Characteristic PBH Halo Mass}

In the Press-Schechter formalism, the characteristic PBH halo mass can be defined as the scale at which the PBH overdensity reaches the critical threshold $\delta_{\rm c}$ giving~\cite{Inman:2019wvr}:
\begin{align}\label{eq:Mstar_full}
    M_{*}(z) & = \frac{2}{\delta_{\rm c}^2} D_{\rm PBH}(z)^2 \fpbh^2 \ m \\ \nonumber
    & \simeq  \frac{2}{\delta_{\rm c}^2} \cdot  \frac{9}{4} \cdot \frac{\Omega_{\rm DM}}{\Omega_{\rm m}} \bigg(\frac{1 + z_{\rm eq}}{1 + z} \bigg)^2 \fpbh^2 \ m
\end{align}
We use this to write Eq.~\eqref{eq:Mstar_main}, with $\fpbh^2 \ m$ replaced by $\langle \fpbh^2 \ m \rangle$, the value averaged over the PBH mass distribution given by:
\begin{align}\label{eq:fsq_mavg}
    \langle \fpbh^2 \ m \rangle & \equiv \fpbh^2  \int m \ \phi(m) \ {\rm d} \ln m.
\end{align}
\subsubsection{PBH Cluster Radius}
In order to estimate the cluster radius at formation, the spherical collapse theory predicts a cluster density given by $\rho_{\rm h} \simeq 178 \cdot \rho_c(z = z_{\rm h})$ 
where $\rho_{\rm h} = 3 M_{\rm h}/(4 \pi r_{\rm h}^3)$. 
This allows us to write cluster radius, after the cluster has virialized \cite{Clesse:2020ghq},
\begin{equation}\label{Eq: Initial M_R}
    r_{\rm h} \sim 70 \ {\rm pc}  \left(\frac{\langle m f_{\rm PBH}\rangle}{\sm} \right)^{1/2}  \left(\frac{M_{\rm h}}{10^6\, M_{\odot}}\right)^{5/6}.
\end{equation}
where we have used Eq.~\eqref{eq:zh} and $\rho_c \propto a^{-3}$.  Note that in Eqs.~\eqref{eq:mh_k}, \eqref{eq:zh}, \eqref{Eq: Initial M_R}, we have corrected some numerical factors found in the expressions of \cite{Clesse:2020ghq}. 

{Using the Press-Schechter formalism, we can additionally compute the fraction of fluctuations that will collapse at redshift $z_{\rm h}$ into PBH halos with a mass $M_{\rm h}$ using:
\begin{equation}
    F(M_{\rm h}, z_{\rm h}) = \mathrm{Erfc} \ \left[\frac{\delta_c}{\sqrt{2} \sigma(M_{\rm h}, z_{\rm h})}\right]
\end{equation}
where the variance $\sigma^2 (M_{\rm h}, z_{\rm h})$ smoothed at the scale $k(M_{\rm h})$ is defined as:
\begin{equation}
    \sigma^2(M_{\rm h}, z_{\rm h}) = \int_0^{\infty} \frac{\mathrm{d}k}{k} \frac{k^3 P_{\rm iso}^{\rm PBH}(z = z_{\rm h})}{2 \pi^2} W^2(k, k(M_{\rm h}))
\end{equation}
with $W(k, k(M_{\rm h}))$ being the top-hat window function defined in Eq.~\eqref{eq:w_k}.  This is valid when 
the Poissonian contribution to the power spectrum dominates over the adiabatic contribution. We find that for $m = 2.6 \ \sm, \fpbh = 1 $, $k(M_{\rm h}) = 400 \ \mathrm{Mpc}^{-1}$, $z_{\rm h} = 30$, $F \sim 0.77$, meaning that already by the $z \sim 30$, more than half of the Poissonian fluctuations at this scale have collapsed to PBH clusters.


\subsection{Dynamical heating of PBH clusters} 

Once a PBH cluster is formed, gravitational interactions between PBHs can lead to a gain in kinetic energy, via a process called dynamical heating, eventually causing the system to ``puff up'' or expand.  This process is similar to the evolution properties of known gravitationally interacting systems, for example,  star clusters.   This effect was used in~\cite{Brandt:2016aco} (see also \cite{Green:2016xgy})} to constrain the abundance of massive compact halo objects (MACHOs) in ultra-faint dwarf galaxies.
In our case, we consider the dynamical heating of PBH clusters, which causes the cluster radius to grow with time (see for example Ref.~\cite{Carr:2023tpt}).
Following \cite{Brandt:2016aco, Carr:2023tpt}, in the case of a monochromatic PBH distribution, the time evolution of the cluster radius is governed by the following equation
\begin{align}\label{eq:R_c_expand}
    \frac{{\rm d} r_{\rm h}}{{\rm d} t} &= \frac{4 \sqrt{2} \ {\pi} G \fpbh m} { 2 v \beta  r_{\rm h}}  \log  \Lambda \\[0.1cm]    \nonumber    
     \mathrm{with} \quad \log  \Lambda & \approx \log \bigg(\frac{r_{\rm h}  v^2}{G \ m} \bigg) \approx \log \bigg(\frac{M_{\rm h}}{2 \ m} \bigg).
\end{align}

We assumed that the typical PBH velocity is of order of the virial velocity, $v \sim v_{\rm vir}$ (see Eq.~\eqref{eq:vvir}).
In Eq.~\eqref{eq:R_c_expand}, $\beta$ is an $\mathcal{O}(1)$ factor that depends on the PBH cluster profile.  We will set $\beta = 3.5 $, which is the expected value for a PBH cluster following a NFW density profile, for consistency with the merger rate calculation.  If one instead assumes a core profile, $\beta$ changes by an order one factor, as well as the final cluster radius $r_{\rm h}$.  Uncertainties on the profile are taken into account in our free clustering parameter $R_{\rm cl}$ (see Sec.~\ref{SubSec: Free Rcl}). 

For a broad mass distribution, (obtained for example for spectral index $n_{\rm s} = 0.965$), additional effects may come from the PBH mass segregation, with heavy PBHs sinking towards the center of the cluster and light PBHs moving towards the periphery.  Numerical N-body simulations would be needed for fully detailed cluster dynamics and are out of the scope of the paper.  We have followed Ref.~\cite{Green:2016xgy} where the dynamical heating by PBHs of stars with a fixed mass in ultra-faint dwarf galaxies was considered for a broad PBH mass distribution and replaced the PBH mass dependent quantity, $f_{\rm PBH} m \ln \Lambda$ in Eq.~\ref{eq:R_c_expand} by the averaged value over the mass function $\langle f_{\rm PBH} m \ln \Lambda \rangle$, given by:
\begin{equation}
    \langle f_{\rm PBH} \ m  \ln \Lambda \rangle = \int m \fpbh(m) \ln \Lambda  \ {\rm d}\ln m
\end{equation}
with $\fpbh(m)$ defined in Eq.~\eqref{Eq: Mass function}. Note that for our final solutions described in sec.~\ref{Sec: PTAs}, we have approximated $\langle f_{\rm PBH} \ m  \ln \Lambda \rangle$ with  $(f_{\rm PBH} \langle m \rangle  \ln \Lambda_{\rm avg})$ with $\Lambda_{\rm avg} = M_{\rm h}/ 2 \langle m \rangle$ and $\langle m \rangle$ being the average PBH mass weighted with the mass distribution (note that this is not the same as $\langle m \rangle$ defined in Eq.~\eqref{eq:m_avg_vask})
\begin{equation}
    \langle m \rangle = \int m \ \phi(m) \ {\rm d}\ln m .
\end{equation}
Our results do not change with this approximation.

In Eq.~\eqref{eq:R_c_expand}, we have neglected the competing cooling effect of any low mass particle like WIMP dark matter or even very light PBHs that could be present in the cluster. {We consistently checked that this effect is indeed systematically smaller than the heating effect. Moreover, the total stellar mass in the typical PBH clusters is expected to be a negligible fraction of the total cluster mass \cite{Simon:2019nxf}. For this reason we also neglect the heating effect induced by the stars.} 

In order to solve Eq.~\eqref{eq:R_c_expand}, we set the initial radius size to the value at cluster formation (after the cluster has virialized), given in Eq.~\eqref{Eq: Initial M_R}. Finally, 
we can write the PBH cluster radius at any time $t$ as:
\begin{align}\label{eq:R_h_heating}
    r_{\rm h}(t)^{3/2} =&r_{\rm h}(t_{\rm c})^{3/2} + \frac{6 \pi}{\beta} \sqrt{\frac{G}{M_{\rm h}}} \langle \fpbh m \ln \Lambda \rangle (t - t_c)  \\ \nonumber
    &
\end{align}
where $t_c$ denotes the time of collapse, at the redshift of cluster formation $z_{\rm h}$. For large $M_{\rm h}$, the first term in the RHS dominates, whereas for small $M_{\rm h}$, the second term from heating will dominate.
%
%
\begin{figure}[t!]
    \centering
    \includegraphics[width=0.5\textwidth]{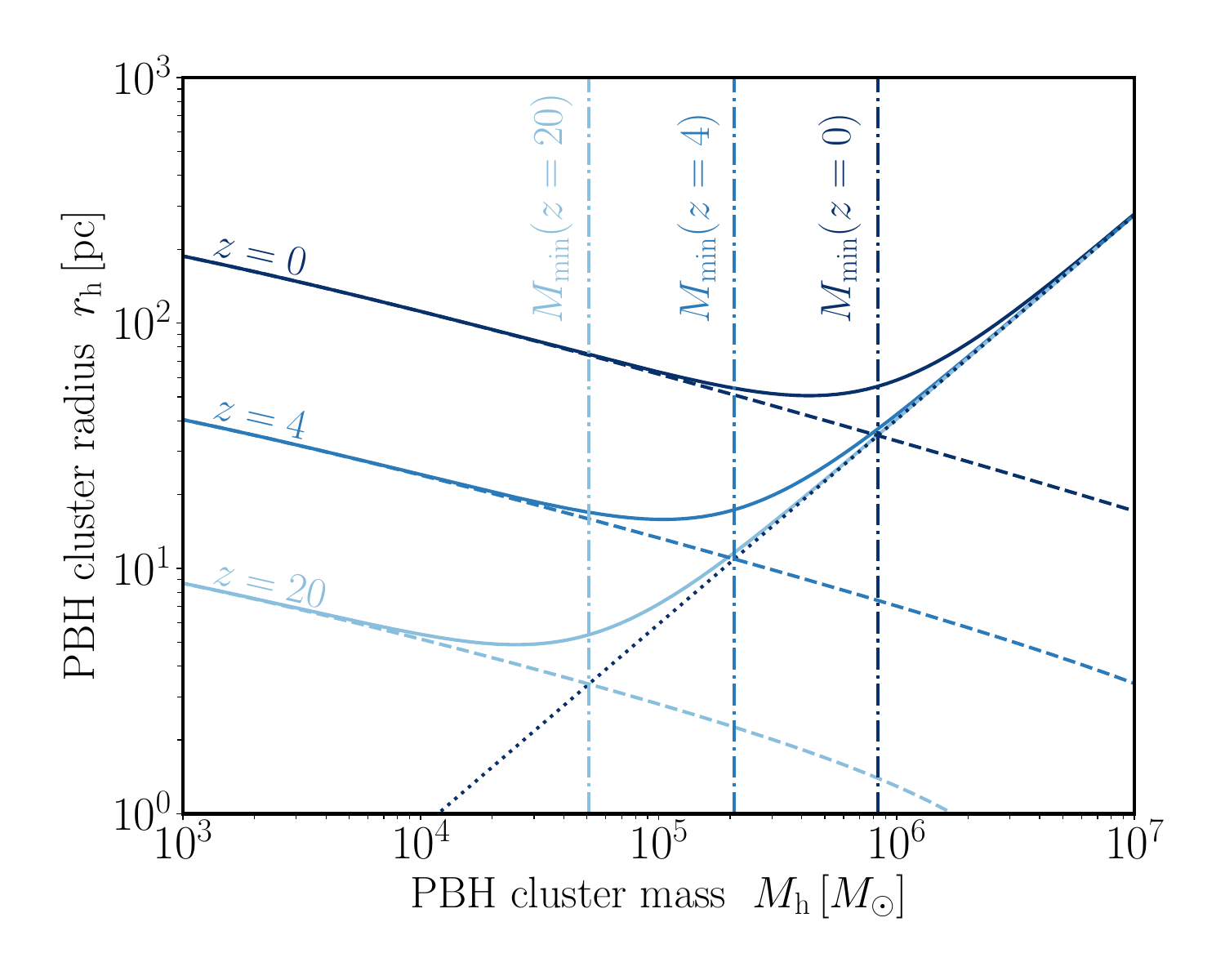}
    \caption{Mass - radius relation given in Eq.~\eqref{eq:R_h_heating} (solid lines) of the PBH clusters at redshift values $z$ fixed to $0, 4, 20$. The large halo masses are unaffected by dynamical heating and follow the mass - radius relation at halo formation (dotted line) (given by Eq.~\eqref{Eq: Initial M_R}). The lighter clusters progressively expand with time and are eventually diluted~\cite{Carr:2023tpt}. The dashed lines show the solution in Eq.~\eqref{eq:R_h_heating} with $r_{\rm h}(t_{\rm c}) = 0$. The minimum cluster mass $M_{\rm min}(z)$ (dotted dashed lines) is given by the intersection of the dotted and dashed lines at the corresponding redshift $z$. In this example, we have fixed $\langle f_{\rm PBH} m \rangle = 3 M_{\odot}$.}
    \label{Fig: Cluster Evolution}
\end{figure}

In Fig.~\ref{Fig: Cluster Evolution}, we show the time evolution of the mass--radius relation for PBH clusters made of PBHs with fixed $\langle f_{\rm PBH} \ m \rangle = 3 \ M_{\odot}$. Initially the clusters follow the initial mass--radius relation given in Eq.~\eqref{Eq: Initial M_R}, with the halo radius growing with the halo mass.  Progressively, the light clusters become affected by the dynamical heating and start expanding. As argued in Ref.~\cite{Carr:2023tpt}, as the cluster expands, it gets disrupted and eventually diluted inside larger clusters. The heating effect is roughly suppressed as $\propto M_{\rm h}^{-1/2}$ for heavier clusters.
In Fig.~\ref{Fig: Cluster Evolution}, we show the minimum cluster mass $M_{\rm min}(z)$ as dashed vertical lines above which the expansion process is not relevant in increasing the cluster size. Numerically, we approximate the minima of Eq.~\eqref{eq:R_h_heating} or $M_{\rm min}$, by finding the halo mass for which the two terms in the R.H.S of Eq.~\eqref{eq:R_h_heating} are equal in magnitude. For those parameter values for which the heating effect is very small i.e. the first term is much larger than the second, we set the mimimum halo mass $M_{\rm min} \sim 10 \  \langle m \rangle$. Additionally, we impose for all scenarios that $M_{\rm min}$ should always be larger than $\sim 10 \ \langle m \rangle$.\\
As time passes, PBH clusters of heavier masses are heated, such that the minimal cluster mass depends on time: $M_{\rm min}(z)$. 


\subsection{Clustering factor $R_{\rm cl}(z)$}
\label{App: Clustering Factor} 
In this section, we provide a calculation of the clustering factor that enters in Eq.~\ref{Eq:Merger Rate}.   We can assemble Eqs.~\eqref{Eq: cross section},\eqref{eq:R_h},\eqref{eq:PS_haloMF} to write:
\begin{align}\label{eq:rcl_defn}
    R_{\rm cl}(z) =  &\frac{4 \pi^2}{3} \bigg(\frac{85 \pi}{6 \sqrt{2}} \bigg)^{2/7} \frac{G^2 \; \bar{\rho}_{\rm DM}^2 }{c^{10/7}}\int_{M_{\rm min}(z)}^{\infty}  \mathrm{d} \, \mh  \\
   & \times \,v_{\rm vir}^{-11/7} \, \mathcal{F_{\rm avg}} \, \mathcal{G}_{\rm NFW} r_{\rm h}^3 \; \delta_{\rm cl}^2 \; \frac{{\rm d} n(z)}{{\rm d} \mh} \; \mathrm{Gpc^{-3} \; yr^{-1}}, \nonumber
\end{align}
where $\mathcal{G}_{\rm NFW}$ is the factor associated to the radial integration of the NFW profile of the halo~\cite{Bird:2016dcv},
\begin{equation}
    \mathcal{G}_{\rm NFW} = \frac{C^3 \, (1-1/\, (1+C)^3)}{9 \, (\, \ln(1+C)-1/(1+C))^2},
\end{equation}
%
while $\delta_{\rm cl}$ parametrizes the overdensity of the halo compared to the background density,
\begin{equation}
    \delta_{\rm cl} = \frac{3 \mh}{4\pi \, r_{\rm h}^3 \, \bar{\rho}_{\rm DM}}.
\end{equation}
Note that here $r_{\rm h}$ corresponds to the cluster Virial radius at the time of cluster formation or collapse given by Eq.~\ref{eq:rcl_defn}.
Using the explicit expression of the halo mass function from Eq.~\eqref{eq:PS_haloMF}, we can further simplify the clustering factor to
\begin{equation}\label{Eq: Integral R_cl}
\begin{split}
     R_{\rm cl}(z) & =   \sqrt{\pi}\bigg(\frac{85 \pi}{6 \sqrt{2}} \bigg)^{2/7} \frac{\left(G \; \Omega_{\rm DM} \rho_{\rm c}\right) ^2 }{c^{10/7}} \int_{M_{\rm min}(z)}^{\infty}  {\rm d}\, \mh   \\
     & \times \, v_{\rm vir}^{-11/7} \, \mathcal{F_{\rm avg}} \,\mathcal{G}_{\rm NFW}\; \delta_{\rm cl} \; \sqrt{\frac{\mh}{M_{*}(z)}} \frac{e^{-\mh/ M_{*}}}{\mh}.
\end{split}
\end{equation}
This approach still contains a certain number of uncertainties and several physical effects may undermine the validity of this calculation, for instance the tidal disruption of clusters in the halo of massive galaxies, collisional disruptions, hierarchical mergers (see Ref.~\cite{Carr:2023tpt} for a discussion on this).  This provides additional motivations to perform an analysis with a free value of $R_{\rm cl}$.

\section{Constraints on the Primordial Power Spectrum }\label{App: Constraints PS}
\subsection{$\mu$-Distortions}
\label{app:firas}
In the analysis performed in Sec.\ref{Sec: PTAs}, we imposed the CMB $\mu$-distortion constraints on the curvature power spectrum at scales $k\lesssim 10^5~\text{Mpc}^{-1}$ \cite{Chluba:2012we}. This is usually parameterized in terms of the parameter $\mu$,
\begin{equation}
    \mu \approx \int^{\infty}_{1 \text{Mpc}^{-1}} \frac{\text{d} k}{k} \,P_{\zeta}(k) W_{\mu}(k),
\end{equation}\label{Eq: mu}
with the window function defined as
\begin{equation}
\begin{split}
    W_{\mu}(k) \approx 2.27 \Bigg[&\exp{\left(-\frac{\left( k/1360\right)^2}{\left(1+\left(k/260\right)^{0.3} +k/340\right)}\right)}\\
    &- \exp{\left(-\left(\frac{k}{32}\right)^2\right)} \Bigg],
\end{split}
\end{equation}
where $\mu$ has been constrained by FIRAS/COBE \cite{Fixsen:1996nj,Mather:1993ij} to have the upper limit $\mu<9\times 10^{-5}$.\\
In our Bayesian inference, we systematically impose this $\mu$-distortion constraint on the power spectrum in Eq.~\eqref{Eq: Primoridal PS}.\\
\subsection{$\Delta N_{\rm eff}$}
Since gravitational waves act as radiation, they are thus constrained by current bounds on the effective number of neutrino species limit, $\Delta N_{\rm eff}  < 0.28$~\cite{Planck:2018vyg}. In our analysis, only the SIGWs will be subject to these constraints since the GWB from late PBH binaries will contribute after CMB. 
This will give the following constrain (see for e.g.~\cite{Dandoy:2023jot}):
\begin{equation}
    \int \text{d}\log(f) \,h^2\Omega_{\rm gw}^{\rm SIGW}(f) < 5.6\times 10^{-6} \Delta N_{\text{eff}},
\end{equation}
where $h^2 \Omega_{\rm gw}^{\rm SIGW}(f)$ is the scalar induced spectrum in the total spectrum defined in Eq.~\eqref{eq:omega_total}.
 The posteriors obtained in Sec.~\ref{Sec: PTAs} satisfy this limit.\\
\subsection{LVK-Constraint}
Finally, the LVK constraint~\cite{KAGRA:2021kbb} (See also Refs.~\cite{Mukherjee:2021ags, Mukherjee:2021itf}) imposes that, at $f_{\rm LVK} = 25 \ \rm Hz$,
\begin{equation}\label{eq:lvk}
    \Omega_{\rm gw} \leq 1.7 \times 10^{-8}.
\end{equation}
The posteriors obtained for the fixed clustering analysis in Fig.~\ref{Fig:fullpos_fixed} do not violate this constraint. For $n_{\rm s} > 1$, $k_{\rm max}$ cuts-off the scalar induced GWB at $f_{\rm max} \sim 10^{-8} \mathrm{Hz}$, much below the LVK frequency range. While for the parameter space with $n_{\rm s} \sim 1$, $\lnAz \lesssim -1.9$ due to the $\fpbh \leq  1$ constraint which gives an $\Omega_{\rm gw}$ satisfying the LVK bound. For the free clustering analysis posteriors for the case with both SIGWs and PBHs  shown in Fig.~\ref{Fig:free_pos_part}, the same reasoning will apply for the parameter space corresponding to $n_{\rm s} \geq 0.8$, where the signal is dominated by the scalar induced GWs. Some parts of the parameter space where $n_{\rm s} \leq 0.8$ corresponding to the GW spectrum originating from PBH binaries can violate the LVK bound. However, the GW background from PBH late binaries for the maximum posterior values satisfies the LVK bound given in Eq.~\eqref{eq:lvk}. 

The best fit values used for the GW spectrum shown in Fig.~\ref{Fig:free_pos_part} are in violation of this LVK bound.

\section{Choice of Window Function}
\label{app:window_func}

Here, we give the formulae for the window function $W(k', k)$ and the linear transfer function $T(k', k)$ used in Eq.~\eqref{Eq: Gaussian Distribution} to define the variance $\sigma_k^2$ of overdensity $\delta$. We want to reiterate here that the relic abundance of PBH exponentially depends on the variance, whose computation relies on the precise shape of the smoothing function $W$. For the choice of $W$, there is currently no prescription (see refs.~\cite{Gow:2020bzo, Young:2019yug, Young:2019osy, Ando:2018qdb} for discussion on this). For our computation of the PBH abundance, we pick the Fourier transform of the real top-hat smoothing function which is one of the common choices used in literature. This window function can be written as:
\begin{equation}\label{eq:w_k}
 W(k', k) = 3 \frac{\sin(\frac{k'}{k}) - (\frac{k'}{k}) \, \cos(\frac{k'}{k})}{(\frac{k'}{k})^3}
\end{equation}
The linear transfer function instead used in Eq.~\ref{eq:sigma_k} is given by~\cite{Young:2019yug}:

\begin{equation}
    T(k', k) = 3 \frac{\sin(\frac{k'}{\sqrt{3} \, k}) - (\frac{k'}{\sqrt{3} \, k}) \, \cos(\frac{k'}{\sqrt{3} \, k})}{(\frac{k'}{\sqrt{3} \, k})^3}
\end{equation}

\section{Full Posterior Distributions}
\label{App: Posteriors}

\begin{figure*}[t!]
    \centering
    \includegraphics[width=14cm]{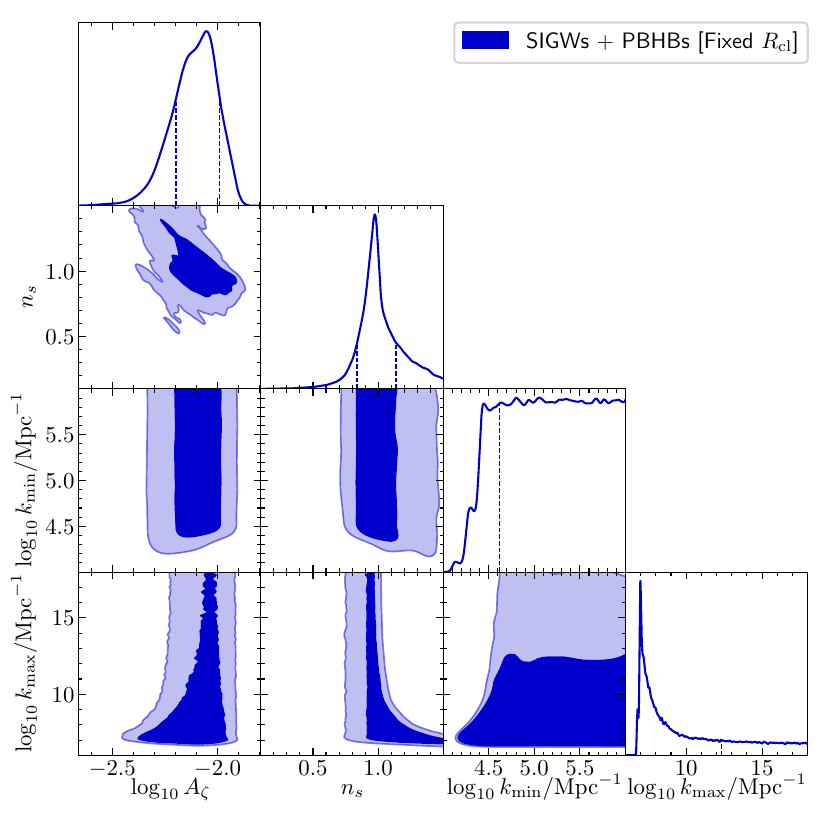}
    \caption{Full posterior distributions for the fixed clustering factor analysis (see Section\ref{SubSec: Fixed Rcl}). The posteriors shown respect $\fpbh \leq 1$ as well as the CMB $\mu$ distortion limits from FIRAS/COBE~\cite{Chluba:2012we}.}
    \label{Fig:fullpos_fixed}
\end{figure*}

\begin{figure*}[t!]
    \centering
    \includegraphics[width=15cm]{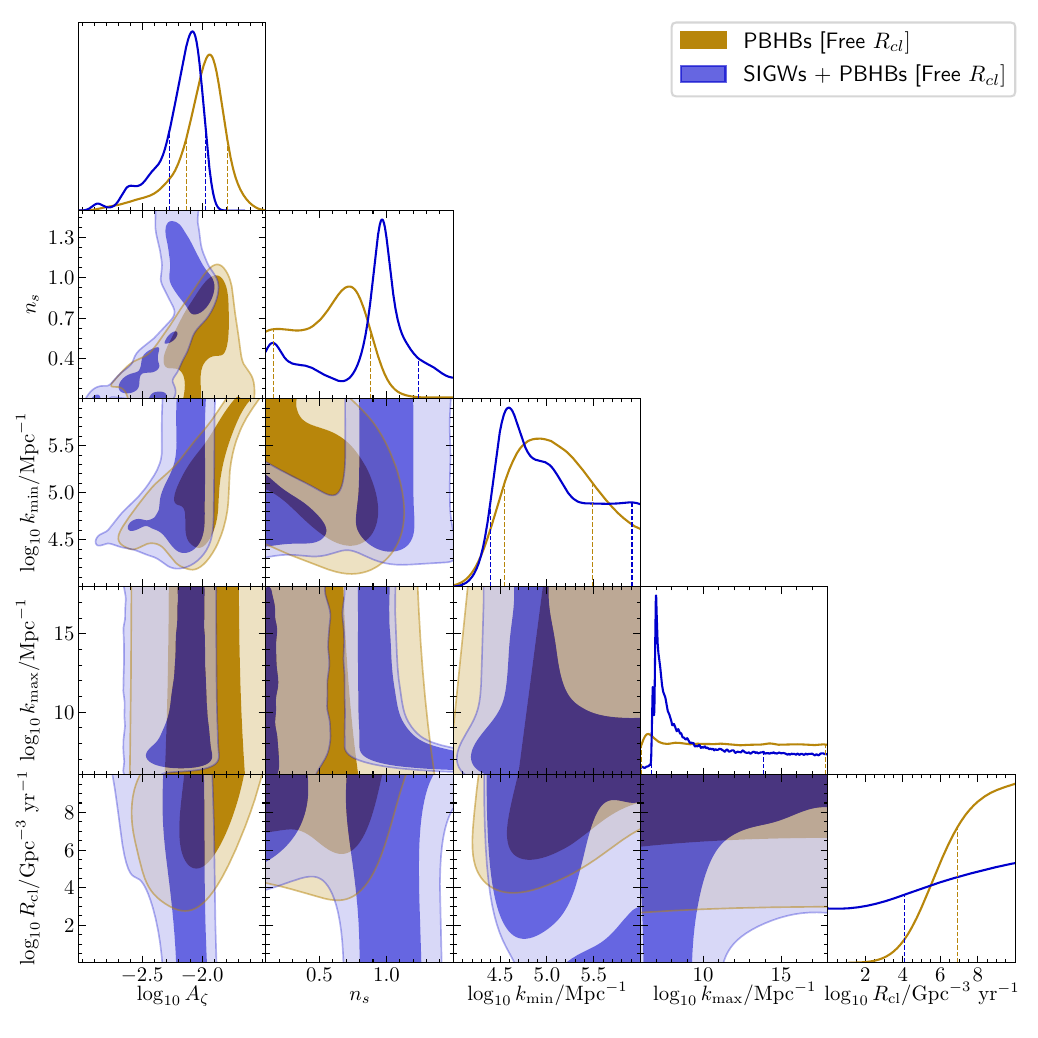}
    \caption{Full posterior distributions for the free clustering factor analysis (see Section~\ref{SubSec: Free Rcl}) for GWB from only PBH binaries (yellow) and GWB including SIGW (blue). The posteriors shown respect $\fpbh \leq 1$ as well as the CMB $\mu$ distortion limits from FIRAS/COBE~\cite{Chluba:2012we}.}
    \label{Fig:fullpos_free}
\end{figure*}

\begin{table*}[t!]
\centering
\begin{tabular}{ |p{1cm}||p{2.5cm}|p{2.5cm}|p{2.5cm}| p{2.5cm}|p{2.5cm}|}
 \hline
 \hline
  \textbf{$\lambda$} & \quad\quad $\log_{10}\, A_{\zeta}$& \quad\quad\quad $n_s$& \quad\quad $\log_{10} \, k_{\rm min}$&  \quad\quad$\log_{10} \, k_{\rm max}$& \quad\quad $\log_{10} \, R_{\rm cl}$\\ [1em]
 \hline
 \hline Prior &\quad\quad$[-4,-1]$& \quad\quad$[0.1,1.5]$& \quad\quad\quad $[4,6]$& \quad\quad $[6,18]$& \quad\quad $[0,10]$\\ [0.5em]
 \hline
\end{tabular}
\caption{Model parameters and their prior range. The parameter $R_{\rm cl}$ is only employed for the free clustering factor analysis, see Section\ref{SubSec: Free Rcl}, and is otherwise computed using App. \ref{App: Clustering Factor}.}
\label{Tab: Prior Distribution}
\end{table*}
In this appendix, we describe the full parameter posterior distributions.

In the right panel of Fig.~\ref{Fig:fullpos_fixed}, we show the full 1d, 2d posterior distributions for the fixed clustering parameter analysis with  $\rcl \equiv \rcl(A_{\zeta}, n_s, k_{\rm min}, k_{\rm max})$. As detailed in the main text, the clustering factor from Poissonian fluctuations is not large enough to enhance the the GW spectrum from PBH binaries. As a result, the 2d posterior region for which the PTA signal can have an important contribution from PBH binaries ($n_s \leq 0.8$) no longer exists. Only the 2d region with $n_s \geq 0.8$ remains, where the PTA GWB can be fitted with only scalar induced GWs (see Fig.~\ref{Fig:fullpos_fixed} 2d panel for $(\log_{10} A_{\zeta}, n_{\rm s})$). 

In the left panel of Fig.~\ref{Fig:fullpos_free}, we show the complete 2d, 1d posterior distributions for free clustering factor analysis (see Section \ref{SubSec: Free Rcl}), including $(k_{\rm min}, k_{\rm max})$. As pointed out in the main text, two regions can be identified in the 2d posteriors for ($A_\zeta$, $n_{\rm s})$ for analysis including both SIGW and PBH binaries (blue): one for which the signal is only fitted by SIGWs and a second for which SIGWs and PBH binaries coexist with similar amplitudes. Looking at the two-dimensional posteriors in $(k_{\rm min}, n_s)$, we indeed clearly see that for $n_s \geq 0.8 $, the distribution becomes degenerated in $k_{\rm min}$. In this parameter space, the signal is fully captured by SIGWs and the GW spectrum is insensitive on $k_{\rm min}$. For $n_s \leq 0.8$ the GW spectrum from PBH binaries contribute to the signal and depends directly on the value of $k_{\rm min}$. The latter acting effectively as a large mass cutoff in the PBH mass function. The same conclusion applies for the posterior distribution in $(R_{\rm cl} ,n_s)$.\\ 
In the same way, the parameter $k_{\rm max}$ acts as a low mass cutoff. In the low spectral index region, $n_s \leq 0.8$ (where the signal is explained by both SIGWs and PBH binaries), the PBH mass function decays quickly at low masses and the parameter $k_{\rm max }$ has little impact on it and therefore on the GW spectrum. For this reason, in this region the distribution is fully degenerated in $k_{\rm max}$ (see $(k_{\rm max}, n_s)$ panel). In the region where the signal is dominated by SIGWs ($n_s>0.8$), and especially when $n_s>1$, the light PBHs become a dominant part of the PBH mass function. Even if the SIGW spectrum is insensitive on $k_{\rm max}$, as $n_s$ increases, the constraint $f_{\rm PBH}\leq 1$ tends to impose a larger value for the cutoff $k_{\rm max}$ to prevent overclosure with light PBHs (see 2d posteriors for $(k_{\rm max}, n_s)$).


\bibliography{references_ms}

\end{document}